\documentclass{aa}  
\usepackage{graphicx}
\usepackage{txfonts}
\usepackage{longtable}
\newcommand{\integral}{{\textit{INTEGRAL~}}}
\newcommand{\lae}{\mathrel{<\kern-1.0em\lower0.9ex\hbox{$\sim$}}}
\newcommand{\gae}{\mathrel{>\kern-1.0em\lower0.9ex\hbox{$\sim$}}}
 
\def\aox{$\alpha_{OX}$}

\def\lx{$L_X$}
\def\lv{$L_V$}

\begin{document}

  \title{The second INTEGRAL AGN catalogue\thanks{All tables of this
  paper are
  also available in electronic form
at the CDS via anonymous ftp to cdsarc.u-strasbg.fr (130.79.128.5)
or via http://cdsweb.u-strasbg.fr/}
}


  \author{V. Beckmann
         \inst{1,2,3},
          S. Soldi\inst{4}, 
          C. Ricci\inst{1,2},
          J. Alfonso-Garz\'on\inst{5},
          T.~J.-L. Courvoisier\inst{1,2},
          A. Domingo\inst{5},
          N. Gehrels\inst{6},
          P. Lubi\'nski\inst{7,1},
          J.~M. Mas-Hesse\inst{5},
          \and
          A.~A. Zdziarski\inst{7}
         }

  \offprints{V. Beckmann}

  \institute{ISDC Data Centre for Astrophysics, Chemin d'\'Ecogia 16, 1290 Versoix, Switzerland\\
\email{beckmann@apc.univ-paris7.fr}
    \and
  Observatoire Astronomique de l'Universit\'e de Gen\`eve, Chemin des Maillettes 51, 1290 Sauverny, Switzerland
    \and
  CSST, University of Maryland Baltimore County, 1000 Hilltop Circle,
  Baltimore, MD 21250, USA
 \and
Laboratoire AIM - CNRS - CEA/DSM - Universit\'e Paris Diderot (UMR 
7158), CEA Saclay, DSM/IRFU/SAp, 91191 Gif-sur-Yvette, France
 \and
Centro de Astrobiolog\'{\i}a LAEX (CSIC-INTA), POB 78, 28691 Villanueva de la Ca\~nada, Madrid,
Spain
    \and
Astrophysics Science Division, NASA Goddard Space Flight Center, Code
661, MD 20771, USA
 \and
Centrum Astronomiczne im. M. Kopernika, Bartycka 18,
PL-00-716 Warszawa, Poland\\
}

  \date{Received 19 March 2009; accepted 30 June 2009}

\abstract
{}
  {The INTEGRAL~mission provides a large data set for studying the
  hard X-ray properties of AGN and allows testing of the
  unified scheme for AGN.}
  {We present analysis of INTEGRAL~IBIS/ISGRI, JEM-X, and OMC
  data for 199~AGN supposedly detected by
  INTEGRAL above 20~keV.}
  {The data
 analysed here allow significant spectral extraction on 148~objects
 and an optical variability study of 57~AGN. 
 The slopes of the hard X-ray spectra of Seyfert~1 and Seyfert~2 galaxies
 are found to be consistent within the uncertainties, whereas
 higher cut-off energies and lower
 luminosities we measured for the more absorbed / type~2 AGN.
 The intermediate Seyfert~1.5~objects exhibit hard X-ray spectra 
 consistent with those of Seyfert~1.  
 When applying a Compton reflection model, the underlying continua appear 
 the same in Seyfert~1 and~2 with $\Gamma \simeq 2$, and the
 reflection strength is about $R \simeq 1$, when assuming
 different 
 inclination angles.
 A significant correlation is found between the hard X-ray and
 optical luminosity and the mass
 of the central black hole in the sense that the more luminous objects appear to be more
 massive. There is also a general trend toward the absorbed sources and
 type~2 AGN having lower Eddington ratios. The black hole mass
 appears to form a fundamental plane together with the optical and
 X-ray luminosity of the form  $L_V \propto L_X^{0.6} M_{BH}^{0.2}$,
 similar to what is found between $L_R$, $L_X$, and $M_{BH}$.}
 {The transition from the type~1 to type~2~AGN appears to be
 smooth. The type~2~AGN are less luminous and have 
 less accreting super massive black holes. The unified~model for Seyfert~ galaxies seems to hold, showing in hard X-rays that the central
 engine is the same in Seyfert~1 and~2, but seen under different
 inclination~angles and absorption. The fundamental~plane links the
accretion mechanism with the bulge of the host galaxy and with the
mass of the central~engine in the same way in all types of Seyfert~galaxies.}

  \keywords{Galaxies: active -- Galaxies: Seyfert  -- X-rays:
              galaxies -- Surveys
              }
  \authorrunning{Beckmann et al.}

  \maketitle

\section{Introduction}
The extragalactic X-ray sky is dominated by active galactic nuclei
(AGN), which are commonly assumed to host an accreting supermassive black hole in
the centres of galaxies. X-ray spectroscopy has been vital in the
study of the AGN phenomenon, because it probes the condition of matter in the
vicinity of the black hole. One model for the X-ray emission is that
of a hot corona lingering on top of the inner accretion disc of the
black hole and emitting inverse Compton radiation from disc photons
that have been upscattered by energetic electrons. 
Another model assumes a disc with a hot inner advection-dominated
accretion flow (ADAF; e.g. \cite{Abramowicz96}). 
An alternative model for the accretion process onto black holes is
that of clumpy accretion flows (e.g. Guilbert \& Rees 1988). 
Courvoisier \& T\"urler (2005) assume that the different elements
(clumps) of the accretion flow have 
velocities that may differ substantially. As a consequence, collisions
between these clumps will appear when the clumps are close to the
central object, resulting in radiation.  

Because optical spectroscopy distinguishes between two main types of low-luminosity
AGN, the broad-line Seyfert~1 and narrow-line Seyfert~2 objects, a
similar distinction is apparent between unabsorbed sources with
on-average softer X-ray spectra and the flatter spectra of absorbed
sources. This has been noticed by Zdziarski et
 al. (1995), based on {\it Ginga} and {\it CGRO}/OSSE data and later
 confirmed e.g. by Gondek et~al. (1996) using combined {\it EXOSAT},
 {\it Ginga}, {\it HEAO-1}, and {\it CGRO}/OSSE spectra, and by Beckmann et~al. (2006) using {\it
   INTEGRAL} IBIS/ISGRI data of AGN above 20~keV. A study of {\it
   BeppoSAX} PDS spectra of 45 Seyfert galaxies has come to
a similar conclusion, although the spectra of Seyfert~2 appeared
steeper when considering a possible cut-off in the spectra of
Seyfert~1 galaxies 
(\cite{Deluit03}).
X-ray data already show that most, but not all, AGN
unabsorbed in the X-rays are Seyfert~1 type, and most, but not all,
AGN that are absorbed belong to the Seyfert~2 group (e.g. Awaki et
al. 1991). 

Thus a longstanding discussion has been, whether these two groups indeed
represent physically different types of objects, or whether they can
be unified under the assumption that they are intrinsically the same
but seen from a different viewing angle with respect to absorbing
material in the vicinity of the central engine (e.g. Antonucci
1993), and that the difference in X-ray spectral slope can be explained
  solely by the absorption and reflection components. This {\it unified model} naturally explains the different
Seyfert types in a way that the broad-line region is either visible
(Seyfert~1) or hidden (Seyfert~2) possibly by the same material in
the line of sight as is responsible for the absorption detectable at soft
X-rays (e.g. Lawrence \& Elvis 1982). 
On the other hand, the model has some problems explaining other aspects of AGN, for example, that some Seyfert galaxies change their type from 1 to 2 and back, but also the observation that Seyfert~2 objects exhibit flatter hard X-ray spectra than Seyfert~1 even in the energy range $>20 \rm \, keV$, where absorption should not play a major role unless  
$N_{\rm H} \gg 10^{24} \rm \, cm^{-2}$. Also, the existence of
Seyfert~2 galaxies that show no absorption in the soft X-rays, like
NGC~3147 and NGC~4698 (\cite{Pappa01}) cannot be explained by the unified model.

Lately, two hard X-ray missions have provided surveys at $> 20 \rm \, keV$
with enough sky coverage to be suitable for population studies of
AGN. One is the NASA-led {\it Swift} mission (\cite{Swift}) launched
in 2004, the other one the ESA-led \integral satellite 
(\cite{INTEGRAL}), launched in October 2002. Due to its observation
strategy of following-up gamma-ray bursts, {\it Swift}/BAT
(\cite{BAT}) provides a more homogeneous sky coverage in the 15--195
keV energy range, while the hard X-ray imager IBIS/ISGRI on-board
\integral is more sensitive and extends up to several hundred keV with
better spectral resolution. \integral provides broad-band coverage
through the additional X-ray monitor \mbox{JEM-X} in the 3--30
keV range (\cite{JEMX}) and provides photometry with the
optical camera OMC in the V-band (\cite{OMC}).

The AGN surveys provided by {\it Swift}/BAT (\cite{Tueller08}) and
\integral IBIS/ISGRI
(\cite{XLF}, \cite{Bassani07}) have already led to the
discovery that the fraction of absorbed and Compton thick sources is
less than expected from cosmic X-ray background synthesis models
(e.g. Treister \& Urry 2005, Gilli et al. 2007). With the ongoing
\integral mission, it is now possible to compile a large sample of AGN
for spectroscopic and correlation studies and to probe the
 unified model for AGN. The data analysis is
described in Sect.~\ref{data_analysis}, the average properties of
the AGN in the sample in Sect.~\ref{properties}, the
discussion of the properties in the view of unified models in
Sect.~\ref{discussion}, and we end with the conclusions in
Sect.~\ref{conclusions}. Notes on individual sources can be found in
the appendix (Sect.~\ref{section:singlesource}).

\section{Data analysis}
\label{data_analysis}

\begin{figure}
\includegraphics[height=8.5cm,angle=90]{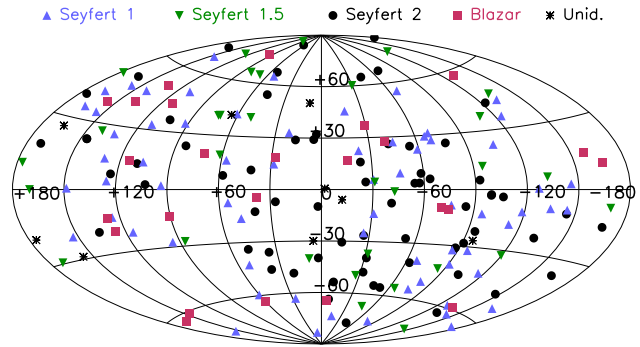}
\caption[]{{\integral}-detected AGN during the first 5 years of the
 mission. As unidentified we mark those sources where the AGN type has not been determined yet.}
\label{fig:skydistribution}
\end{figure}

The list of AGN presented here is based on all \integral
detections of AGN reported in the literature, therefore
enter into the \integral general reference catalogue\footnote{for
 the latest version of the catalogue see http://isdc.unige.ch/index.cgi?Data+catalogs}
(\cite{refcat}, Bodaghee et~al. 2007). It has to be pointed out that
 for many sources, we present the first \integral spectral analysis,
 because Bodaghee et~al. (2007), Sazonov et~al. (2007), and Bassani et~al. (2006) did not include spectral analysis, and Beckmann et~al. (2006)
 discussed a sample of 38 AGN based only on 1.3 years
 of INTEGRAL data. With the \integral mission
continuing smoothly, most of the sky has been observed in the first 5
years of operations, leading to a rather uniform sky distribution of
detected AGN, as shown in Fig.~\ref{fig:skydistribution}. 
For each extragalactic source,
we analysed the IBIS/ISGRI, JEM-X, and OMC data from the early mission
(revolution 26 starting on 30~December 2002) up to spacecraft
revolution 530 (ending on 17~February 2007), covering more than 4
years of data. To include
only high-quality data, the selection considers ISGRI data taken at an
off-axis angle smaller than $10^\circ$ and includes only
those observations that lasted for at least 500 sec. Analysis
software used in this work is version 7 of the Offline Standard
Analysis Software (OSA) provided by the ISDC Data Centre for
Astrophysics (\cite{ISDC}). For each source,
imaging analysis was performed to determine the significant
sources in the field around the AGN. Taking their fluxes into account is
important when analysing data from coded-mask instruments, because all
sources in the field add to the background of the source of interest. Then standard spectral extraction
was used, considering all significant sources in the field. 

For the X-ray monitor JEM-X, a similar selection of data was performed, using a maximum off-axis angle of $3^\circ$ due to the
smaller field of view compared to IBIS. The much lower effective exposure time when
compared to IBIS/ISGRI (Table~\ref{catalog}) results in only 23
detections of AGN by JEM-X with a significance $> 5\sigma$.
For JEM-X, spectra were extracted from
the mosaic images, since this procedure is more reliable for faint
sources than the standard spectral extraction in OSA~7. 

Naturally, some sources reported in the literature do not show up
significantly in the data analysed here, because they were observed
after February 2007, or they are, like the blazar class, highly
variable and therefore do not give a significant detection in the
combined data set. The $199$ AGN reported to be
found in {\integral} data are listed in Table~\ref{catalog},
together with their redshift, position (J2000.0), and their effective
exposure time in IBIS/ISGRI and JEM-X for the data set used here.  
Twelve sources, which were reported in the literature but gave a
  detection significance $< 3 \sigma$ in the data presented here are listed in
  Table~\ref{catalog} and marked by an {\it x}. These objects are not considered in the following
  analysis.
All errors given in this paper are at the $1 \sigma$ level.

\subsection{Black hole masses}
\label{bhmasses}

We also include in Table~\ref{catalog} the black hole masses of the central
engine and the method
used to determine them, as found in the literature.
Different methods can be used to estimate the mass of the central black hole $M_{\rm BH}$
in an AGN or a normal galaxy, most of them still carrying fairly large uncertainties.
Nevertheless, considering the importance of the black hole mass in studying the properties of these objects,
we decided to include a compilation of the mass estimates from the literature as the best guess that can be provided 
at present for each object in this catalogue.

We have included masses estimated from gas and/or stellar kinematics in the nuclear region of the galaxy,
in the presence (method `M', see e.g. \cite{Greenhill97}) or not (`K', \cite{Hicks08}) of a water maser,
from assuming virialized motions of the broad line region (BLR) clouds, either using the reverberation-mapping 
technique (`R', \cite{Kaspi00}) or estimating the size of the BLR from the emission line luminosity
(of the $H\beta$ line usually; `LL', \cite{Wu04}) or from the optical continuum luminosity (usually
measured at 5100 \AA; `CL', \cite{Kaspi00}). Other methods are based
on the empirical relation between the black hole mass and 
the stellar velocity dispersion $\sigma_s$, using either direct measurements of the latter 
(`S', \cite{Ferrarese00}) or indirect estimates of $\sigma_s$ from the width of the [O III] line
(`SO', \cite{Greene05}) or from the morphological parameters of
the bulge (`SB', \cite{ODowd02}). Some estimates use the bulge luminosity (`B', \cite{Wandel02}), the K-band stellar magnitude (assuming that it is 
dominated by the bulge; `KM', \cite{Novak06}), the X-ray variability
time scales (`X', \cite{Gierlinski08}), or the properties of outflowing warm absorber clouds (`W', \cite{Morales02}).
Whenever the uncertainty on the estimate of the black hole mass is not available
in the reference paper, we assumed a conservative one following the typical uncertainties
of the method used for the mass measurement.

The most reliable methods are those involving direct measurements of gas and stellar kinematics, with average uncertainties
in the range 0.15--0.3 dex, reaching 0.1 dex or less when water maser emission is detected (see \cite{Vestergaard04} for more details).
Also the reverberation mapping technique provides black hole masses with accuracy around 0.15--0.3 dex, which 
drops to values of 0.4--0.5 dex and even to 1 dex when the radius of the BLR is estimated from the emission line or the continuum luminosity.
Masses estimated from the stellar velocity dispersion can have uncertainties around 0.3 dex when $\sigma_s$ is directly
measured, while indirect measurements of $\sigma_s$ result in much less precise estimates ($\ge 0.7 \rm \, dex$).
Larger uncertainties are provided by the other methods mentioned above, 0.5--0.6 dex for method `B', 0.5--1 dex for `X' (\cite{Awaki05}),
0.5 dex for `KM' (\cite{Winter09}), and only upper limits can be derived with the method based on outflowing warm absorber clouds.

\subsection{X-ray spectral fitting}
For all $187$ objects with a detection significance above $3 \sigma$ in the IBIS/ISGRI
18--60~keV energy band, spectral analysis was performed using an
absorbed power law with $N_H$ fixed to the value reported in the
literature (Table~\ref{fitresults}) and adopting XSPEC version 11.3.2
(\cite{XSPEC}). When the significance was below $5 \sigma$, the photon
index was fixed to $\Gamma = 2$. The $N_H$ value used for the fitting is the intrinsic absorption plus the Galactic
hydrogen column density, whereas in Table~\ref{fitresults} only the intrinsic
absorption is reported.
In
cases where no absorption information was found, {\it Swift/XRT} and
{\it XMM-Newton} data were analysed to determine the
level of absorption. For those
objects detected only at a low significance level
(i.e. between 3 and $5\sigma$), the photon index was fixed to $\Gamma = 2.0$ in order to extract a flux value. 
Table~\ref{fitresults} gives
the fit results to the IBIS/ISGRI data.  
Fluxes are model fluxes according to the best-fit result. In
the cases where a cut-off power law model gave a significantly
better fit to the ISGRI data we set the $\Gamma$ column to `C'. For
these 12 objects, the fluxes reported are based on the best-fit model
reported in Table~\ref{complexISGRIspec}. These 12 sources and
  all sources that gave a high IBIS/ISGRI detection significance of
  $>30 \sigma$ are also discussed in more detail in Appendix~\ref{section:singlesource}.

The JEM-X
spectra of the 23 AGN detected by the X-ray monitor were fit with the IBIS/ISGRI data and results reported in Table~\ref{fitcombined}. As for the ISGRI spectra alone, we also did not fit the absorption
values in the case of the combined JEM-X/ISGRI spectra, because the
JEM-X data starting at $3 \rm \, keV$ did not allow a significant
constraint on $N_{\rm H}$ in most cases. In cases
where the flux of the source varied significantly, so no combined
fit could be performed resulting in $\chi^2_\nu < 2$, only
simultaneous data were used (e.g. in the case of NGC~4388). For
two AGN, NGC~1275 and IGR~J17488--3253, a more complex model than an
absorbed cut-off power law was required to represent the combined JEM-X and IBIS/ISGRI data (see Appendix~\ref{section:singlesource}).

\subsection{Optical data}
Optical data in the V band are provided by the optical monitoring camera (OMC).
Data were extracted from the OMC Archive\footnote{http://sdc.laeff.inta.es/omc/}
getting one photometric point per shot. The photometric apertures were centred
on the source position, as listed in version 5 of the OMC Input Catalogue
(\cite{Domingo03}). The fluxes and magnitudes were derived from a photometric
aperture of $3\times3$ pixels (1 pixel = 17.504 arcsec), slightly circularized,
i.e. removing $\frac{1}{4}$ pixel from each corner (standard output from OSA).
Therefore the computed values include the contributions by any other source
inside the photometric aperture. We flagged in Table~\ref{OMCresults} those
sources that might be affected by a nearby star (at less than 1\arcmin), with a
potential contamination below 0.2 mag in any case. Other 8 AGN containing a
brighter contaminating source within the extraction aperture were not included
in this compilation.  In addition, for some extended AGN, this $3\times3$
aperture does not cover the full galaxy size, but just their central region.

To only include high-quality data, some selection criteria
were applied to individual photometric points. Shots were checked against saturation,
rejecting those with long exposures for the brightest sources, if
necessary. For faint sources, a minimum signal-to-noise ratio of 3 was required for
the longest integration shots. The shortest shots were only used if the signal-to-noise
ratio was greater than 10. Because these sources can show extended
structure in the OMC images, anomalous PSF, as well as problems in the centroid
determination, were allowed. Finally, to avoid contamination by cosmic rays, we excluded
those points whose fluxes deviate more than 5 times the standard
deviation from the median value of their surrounding points, applying three
iterations of this filter. 

We list in Table~\ref{OMCresults} the median V magnitude of each AGN, the
average of error estimates ($1\sigma$ level) of each photometric point given by
OSA~7, $\langle \sigma_V \rangle$, the luminosity in the Johnson V filter (centred on 5500
\AA, effective width 890 \AA), the \aox\ value, the number of photometric points
used in the analysis and a flag indicating the potential contamination of the
photometric value by a nearby star. The value of $\alpha_{OX}$ is measured as the slope of a
power law between the two energy ranges 
\begin{equation}
\alpha_{OX} = - \frac{\log (f_O / f_X)}{\log (\nu_O / \nu_X)} .  
\end{equation}
Here, $f_O$ and $f_X$ are the monochromatic fluxes at the frequencies $\nu_O$ (at
5500 \AA) and $\nu_X$ (at 20 keV). 

No $K$ correction has been applied to the V luminosities, since the redshifts
are relatively low and the optical slope of these objects is not well known.
Moreover, depending on the redshift, the V band might be contaminated to
different degrees by the OIII and H$\beta$ emission lines.  We did not
correct for this effect, either. 

\subsubsection{Optical variability}
\label{opticalvariability}

Among the 57 AGN for which OMC data are available, 3 show strong variability in
the photometric V-band data, with an amplitude larger than 0.5~mag: QSO~B0716+714, NGC~4151, and 3C~279.   

\begin{figure}
\includegraphics[width=8.5cm]{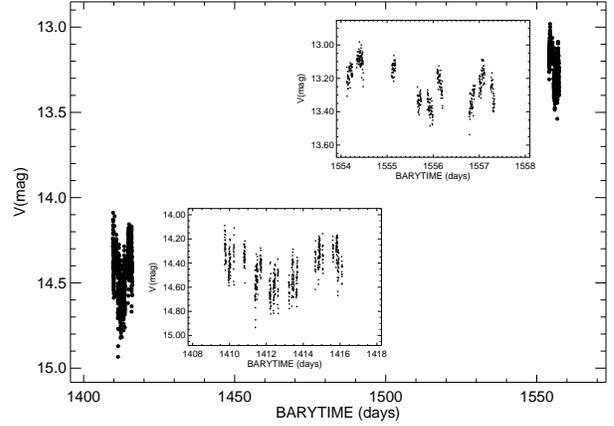}
\caption[]{OMC lightcurve of QSO~B0716+714. The insets show a zoom on the
lightcurves during the 2 periods when this object was monitored by OMC. The
optical luminosity increased by a factor of 4. The photometric accuracy of
individual photometric points is $\sigma \sim0.09$~mag. Barycentric time is
given in \textit{INTEGRAL} Julian Date 
(IJD).}
\label{fig:QSO-OMC}
\end{figure}

The BL~Lac QSO~B0716+714 appears as a point source in a low background
field and was
monitored by OMC during 2 periods, at IJD\footnote{The INTEGRAL Julian
  Date is defined as IJD$=$MJD$-$51\,544.0. The origin of IJD is 2000 January 1 expressed in Terrestrial Time.} around 1415 (November 2003) and 1555
(April 2004). This source brightened by a factor $\sim 4$ during this period, as
shown in Fig.~\ref{fig:QSO-OMC}, ranging from $ V = 14.60 \rm \,
  mag$ to 13.05~mag. Moreover, this blazar also shows a day-scale
variability pattern within the two monitoring periods, with an amplitude around
0.3~mag.

NGC~4151 is an extended source, classified as Seyfert~1.5, much larger than
the OMC aperture. We detected a clear weakening of its central,
dominating region by around 0.5 mag (60\% in flux) from 11.15~mag to 11.65~mag between May 2003 and
January 2007. Since the optical photometry is contaminated to some extent by bulge stellar light, 
the variation in the optical emission from the nucleus itself 
might
have been significantly larger. 

\begin{figure}
\includegraphics[width=8.5cm]{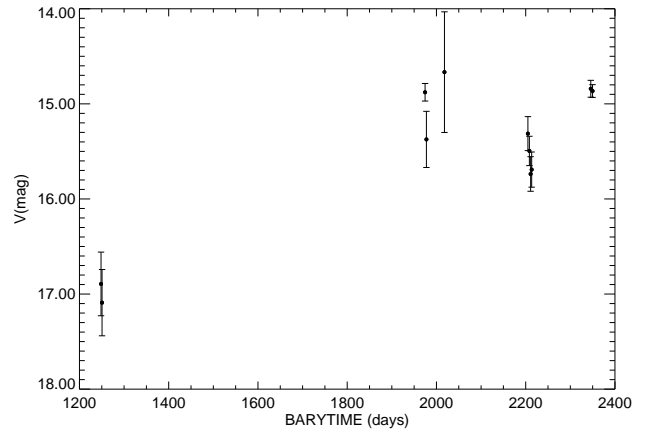}
\caption[]{OMC lightcurve of 3C~279. Barycentric time is given in IJD.
Photometric data have been re-binned to obtain one
value per revolution (3 days).}

\label{fig:3C279-OMC}
\end{figure}

The AGN 3C~279 was barely detectable around June 2003, with a $2\sigma$ OMC
detection at $V = 17.0$. The catalogued value for it in the low state is
$V = 17.8$ mag (\cite{ODell}). After May 2005, this blazar was clearly detected
with a brightness in the range $V = 14.8-15.8$ mag, indicating that it
brightened by up to a factor close to 10 (Fig.~\ref{fig:3C279-OMC}). This is a very active source, with optical
photometry reported in the range $B = 18.3$ to $B = 11.3$ (\cite{ODell}).

Other AGN monitored by OMC also show some hints of variability, such as 3C~273 and
3C~390.3,  but at smaller amplitudes (just a few percent over the period
considered).  


\section{Properties of INTEGRAL detected AGN}
\label{properties}

The catalogue of {\integral}-detected extragalactic objects presented here comprises
187 sources in total. Out of this sample, 162 objects have been
identified as Seyfert galaxies (161 with redshift information), 18
blazars (all of them with redshift), 
and 7 objects have
been claimed to be AGN without further specification of the AGN
type. Within the Seyfert group, we found 67 Seyfert 1 to 1.2 objects,
29 intermediate Seyfert 1.5, and 66 Seyfert type 1.9 and type 2.

\begin{figure}
\includegraphics[height=8.5cm,angle=90]{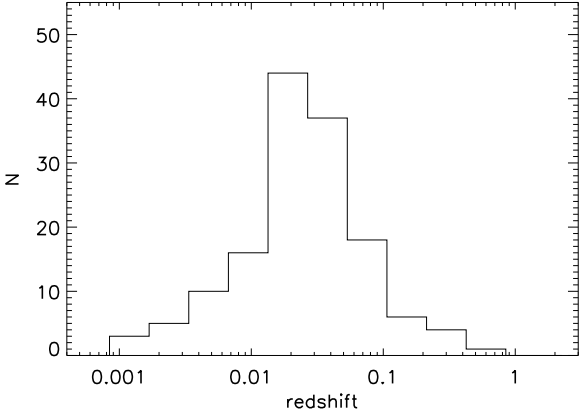}
\caption[]{Redshift distribution of the Seyfert content of 144 {\it
   INTEGRAL} detected AGN with detection significance $\ge
   4\sigma$. The average redshift is $z = 0.03$.}

\label{fig:zdistribution}
\end{figure}

\begin{figure}
\includegraphics[height=8.5cm,angle=90]{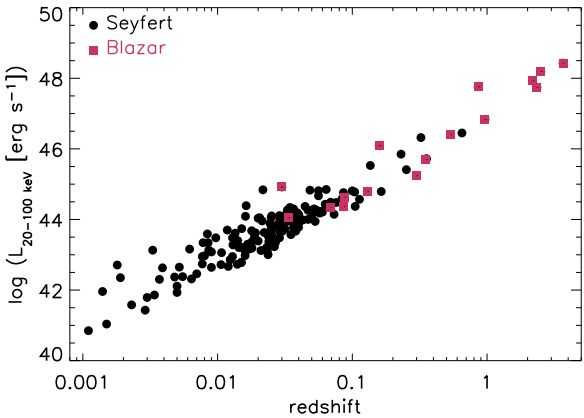}
\caption[]{X-ray luminosity versus redshift of 161 {\it
   INTEGRAL} detected AGN with significance $\ge 4\sigma$.}

\label{fig:z_Lx}
\end{figure}

The redshift distribution of the Seyfert type AGN in this sample is
shown in Figure~\ref{fig:zdistribution}. The average redshift is $z = 0.03$. 
Figure~\ref{fig:z_Lx} shows the parameter space filled by  {\it
 INTEGRAL}-detected AGN in redshift and X-ray luminosity, ranging
 from low-luminous low-redshift Seyfert as close as $z=0.001$ up to the
 high-redshift blazar domain. 

To investigate the spectra of AGN subtypes, we derived
averaged spectral properties and stacked spectra of the Seyfert 1 and 2 types, as well as for the
intermediate Seyferts and the blazars, and according to the intrinsic
absorption. 
The
nine brightest sources, two blazars and seven Seyfert, with $>50 \sigma$ 
(Cen~A, NGC~4151, GRS 1734-292, NGC 4945, NGC 4388, IC 4329A, Circinus
Galaxy, 3C 273, and Mrk 421)
have been excluded from the statistical spectral
analyses because their high signal-to-noise ratio would dominate the
averaged spectra, and also all sources with $< 5\sigma$ were ignored. In addition, we excluded the 5 Compton thick objects,
i.e. NGC~1068, NGC~3281, NGC~4945, Circinus~Galaxy, ESO~138-1.

The average Seyfert 1 (including type 1.2) spectral property was constructed
using the mean weighted by the errors on the photon indices
of 55 ISGRI power-law fit results, the Seyfert 2 composite
spectrum includes 44
sources, and 20 objects form the intermediate Seyfert 1.5 group where
spectral fitting allowed constraining the spectral shape (Table~\ref{fitresults}). In
addition, 11 blazars 
allowed spectral extraction. 
When computing the weighted average of the various
subclasses, the 11 blazars had a hard X-ray spectrum with $\Gamma = 1.55 \pm
0.04$ when compared to the 119 Seyfert galaxies with $\Gamma = 1.93 \pm 0.01$. The Seyfert 1 ($\Gamma = 1.92 \pm 0.02$) and Seyfert 1.5
($\Gamma = 2.02 \pm 0.03$) only show slightly steeper hard X-ray spectra than the Seyfert 2 objects ($\Gamma =
1.88 \pm 0.02$). Table~\ref{average} gives the properties of the different
Seyfert types. All quantities, except for the photon indices, have
been averaged in logarithmic space. For 12 objects, a cut-off
  power law model gave a better representation of the ISGRI
  spectra (Table~\ref{complexISGRIspec}). The average photon index is in
  these cases $\langle \Gamma \rangle = 1.3 \pm 0.4$ with a cut-off energy of  $\langle E_C \rangle = 86 \pm 25 \rm \, keV$.

We get a similar result when stacking the IBIS/ISGRI spectra
together. Again, only sources above $5
  \sigma$ are considered here.
The spectra were renormalised on the 18--30 keV
energy bin before stacking them, re-adjusting the errors on the flux so
that the significance is taken into account. The significances for
the stacked spectra are consistent with what would be expected based
on the single spectra significances, i.e. the 18--60 keV significances
are $98 \sigma$ for the Seyfert~1, $61 \sigma$ (Seyfert~1.5), $94
\sigma$ (Seyfert~2), and $148 \sigma$ for all Seyfert spectra stacked
together. The results of spectral model fitting to these spectra are
summarized in Table~\ref{spectralfits}. A simple power law model
gives a photon index of $\Gamma = 1.97 \pm 0.02$ for all Seyfert
objects, $\Gamma = 1.96 {+0.03 \atop -0.02}$ for the Seyfert~1,
$\Gamma = 2.02 \pm 0.04$ for Seyfert~1.5, and $\Gamma = 1.89 {+0.04 \atop -0.02}$
for the Seyfert~2 class. In all cases, a cut-off power law improves
the fit result significantly according to an F-test
(Table~\ref{spectralfits}). The resulting model is $\Gamma = 1.4 \pm
0.1$ with cut-off at $E_C = 86 {+16 \atop -17} \rm \, keV$ for all
Seyfert galaxies, fully consistent with the average from the 12 
  cut-off power law model fits. For Seyfert~1 we derive $\Gamma = 1.4  \pm
0.1$ and $E_C = 86 {+21 \atop -14} \rm \, keV$, and $\Gamma = 1.4 \pm
0.2$ with $E_C =  63 {+20  \atop -12} \rm \, keV$ for intermediate
Seyfert~1.5, and $\Gamma = 1.65 \pm
0.05$ with $E_C = 184 {+16 \atop -52} \rm \, keV$
Seyfert~2 galaxies. It should be considered that
  the latter fit still has a poor quality with $\chi^2_\nu = 2.6$.  

\begin{figure}
\includegraphics[height=8.5cm,angle=270]{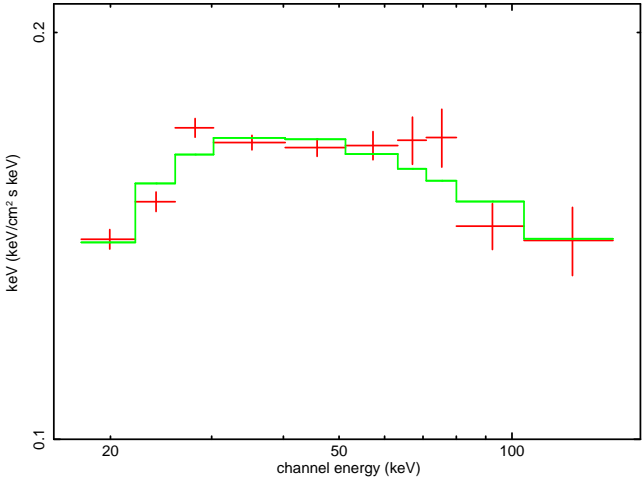}
\caption[]{Stacked IBIS/ISGRI spectrum for Seyfert
 objects (excluding Compton thick sources) in $E F_E$ versus
 energy. The spectrum has been fit by a PEXRAV model 
with $\Gamma = 1.95$ and reflection $R = 1.3$.}

\label{fig:stackSy}
\end{figure}

A model that adds a reflection component from cold material to the
underlying continuum (the so-called PEXRAV model; Magdziarz \&
Zdziarski 1995) gives again a better fit
in most cases. The
underlying continuum shows a similar gradient in the
different source classes when not allowing for a high-energy cut-off
in the PEXRAV model, so this fit has the same degree
  of freedom as the one using the cut-off power law. The difference
shows up, however, in the inclination angle $i$ and strength $R$ of the
reflection component. Here $R$ is defined
as the relative amount of reflection compared to the directly viewed
primary spectrum. The value of $R$ depends on the
inclination angle $i$ between the normal of the accretion disc and
the line of sight. The smaller the inclination angle, the larger the
resulting reflection component. As the data are not sufficient to fit
$R$ and $i$ simultaneously, the inclination angle was set to $i = 30^\circ$ for
Seyfert~1, $i = 45^\circ$ for Seyfert~1.5, and $i = 60^\circ$ for
Seyfert~2. It is worth noting that the quality of the fit did not
depend on the choice of $i$. The results are included in
Table~\ref{spectralfits}. Using this model,
Seyfert~1 and Seyfert~2 show only slightly different underlying continua and
a reflection component of the same strength $R \simeq 1$ within the
statistical errors. Seyfert~1.5 objects appear to have slightly
steeper spectra ($\Gamma = 2.0$) and stronger reflection ($R = 3 {+5
 \atop -1}$). Applying a higher significance level (e.g. $\ge 10 \sigma$)
for the source selection does not change the results
significantly, as shown for all Seyfert galaxies in Table~\ref{spectralfits}. This confirms that the stacked spectra are dominated
by the most significant sources.
The result of the fit to the whole Seyfert sample with the PEXRAV model
is shown in Fig.~\ref{fig:stackSy}.

The classification according to the Seyfert type of the objects is
based on optical observations. An approach to classifying sources due to
their properties in the X-rays can be done by separating the sources
with high intrinsic absorption ($N_{\rm H} > 10^{22} \rm \, cm^{-2}$)
from those objects that do not show significant absorption in the
soft X-rays. Not all objects that show
high intrinsic absorption in the X-rays are classified as Seyfert~2
galaxies in the optical, and the same applies to the other AGN
sub-types. Nevertheless a similar trend in the spectral slopes can be
seen: the 44 absorbed AGN show a 
hard X-ray spectrum ($\langle\Gamma \rangle =1.91 \pm 0.02$) consistent
with that of 
the 66 unabsorbed sources ($\langle \Gamma \rangle
=1.94 \pm 0.02$). 
Using the stacked spectra, the absorbed sources show a slightly flatter continuum
with $\Gamma = 1.91 {+0.04 \atop -0.03}$ than unabsorbed Seyfert galaxies, with 
$\Gamma = 1.97 {+0.03 \atop -0.01}$. Also here, a cut-off power law
  has been tested, but comparing it to the simple power law model improves the fit only for the unabsorbed
  sources. For these sources we derive $\Gamma = 1.5  \pm
0.1$ and $E_C = 100 {+25 \atop -15} \rm \, keV$, and for the absorbed
ones $\Gamma = 1.4 \pm
0.1$ with $E_C =  94 {+32  \atop -13} \rm \, keV$. Thus, for the
different absorption classes we get consistent cut-off values and
spectral slopes, although the stacked unabsorbed spectrum is
represented better by the simple power-law model with no cut-off.
Applying the PEXRAV model shows that the spectra
can be represented by reflection models. The fit improves
significantly when adding a reflection component with $R \simeq 1.5$, while
the underlying continuum slope is the same as for the simple power
law.

A difference between type 1 and type 2 objects is seen in the
average luminosity of these subclasses. For 60 absorbed Seyfert
galaxies, the average luminosity is $\langle L_{20 - 100 \rm keV} \rangle = 2.5
\times 10^{43} \rm \, erg \, s^{-1}$, more than a factor of 2
lower than for the 74 unabsorbed Seyfert with redshift information ($\langle L_{20 - 100
 \rm keV}\rangle  = 6.3 
\times 10^{43} \rm \, erg \, s^{-1}$). 
The differences in luminosities are exactly the
 same when excluding the 5 Compton thick objects. The 16 blazars again
 appear brighter
when assuming an isotropic emission, with 
$\langle L_{20 - 100 \rm keV}\rangle  = 10^{46} \rm \, erg \, s^{-1}$. 
The latter value has to be used
with caution: because the blazar emission is beamed towards the observer,
not isotropic but collimated in a jet, blazars are highly variable
and {\integral} detects them mainly in phases of outbursts.

\section{Discussion}
\label{discussion}

\subsection{The sample in comparison with previous studies}
\label{comparisonprevious}

\begin{figure}
\includegraphics[height=8.5cm,angle=90]{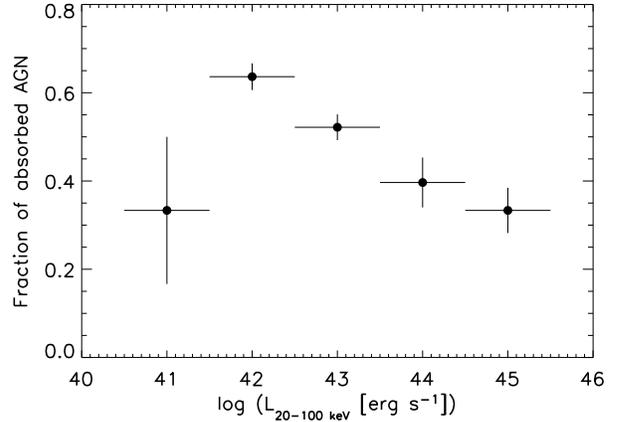}
\caption[]{Fraction of absorbed ($N_{\rm H} > 10^{22} \rm \, cm^{-2}$)
 Seyfert galaxies as a function of hard X-ray luminosity for {\it
   INTEGRAL} detected AGN ($\ge 4 \sigma$). The lowest
   luminosity bin includes only 3 sources.}
\label{fig:NH_Lx}
\end{figure}

For the whole population of sources seen by \integral we observe an
increase in the fraction of unabsorbed objects compared to the
first \integral AGN catalogue. Whereas $\frac{2}{3}$ of the Seyfert population in the first
sample showed $N_{\rm H} > 10^{22} \rm \, cm^{-2}$, there are now
 more unabsorbed than absorbed sources, i.e. only 44\% appear to be absorbed. A similar trend has also been
observed in the {\it Swift}/BAT survey (\cite{Tueller08}), where the
fraction is $\frac{1}{2}$. 
This trend is expected because with ongoing observations, IBIS/ISGRI and
BAT perform deeper studies, which is also reflected in the increase in the
average redshift and luminosity of the objects that are detectable. Sazonov et
al. (2004, 2007) have shown that
there is an anticorrelation of the fraction of absorbed sources with
the X-ray luminosity. In a sample of 95 AGN detected by {\it RXTE}, they
observed that the fraction of absorbed sources is $\sim 70\%$ for
$L_{3 - 20 \rm \, keV} < 10^{43.5} \rm \, erg \, s^{-1}$ but only
$\sim 20 \%$ for objects with $L_X > 10^{43.5} \rm \, erg \, s^{-1}$.
This trend can also be seen in the sample presented here. Determining
the fraction of absorbed Seyfert galaxies in logarithmic bins of X-ray
luminosity, we see the trend toward a decreasing fraction with luminosity
in the range $L_{20-100 \rm \, keV} > 3 \times 10^{41}\rm \, erg
\, s^{-1}$ (Fig.~\ref{fig:NH_Lx}). The fraction appears low though in
the lowest luminosity bin ($L_{20-100 \rm \, keV} < 3 \times 10^{41}\rm \, erg
\, s^{-1}$), but it has to be considered that this luminosity
bin contains only 3 AGN, the Seyfert 1.5 galaxies NGC~4258, NGC~4395, and NGC~5033.

The fraction of Compton thick objects in the sample presented here is only $4 \%$ (5 objects out of the 135
Seyfert galaxies with measured intrinsic absorption). Although this sample is not a complete one,
this indicates further that the fraction of Compton thick AGN is
indeed $\ll 10\%$ as already reported in e.g. Beckmann et al. (2006b)
and Bassani et al. (2007). In addition, a recent study based on combined
\integral and {\it Swift}/BAT data puts an upper limit of $\lae 9\%$ on the
fraction of Compton thick AGN (\cite{Treister09}), and one can thus consider most of the Seyfert population detected
by \integral and {\it Swift} to be Compton thin. 
\begin{figure}
\includegraphics[height=8.5cm,angle=90]{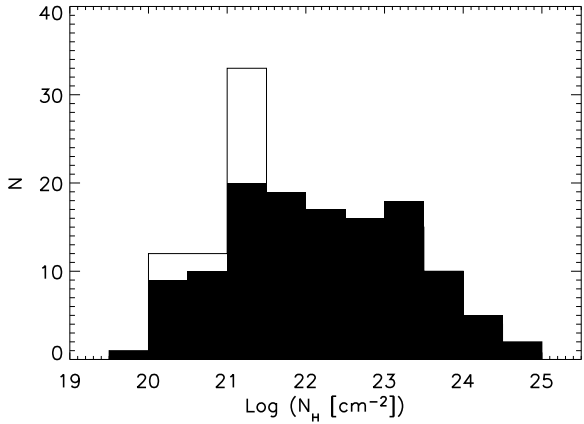}
\caption[]{Distribution of intrinsic absorption of the Seyfert content of {\it
   INTEGRAL} detected AGN ($\ge 4 \sigma$). The unshaded area indicates objects
   where the $N_{\rm H}$ value is an upper limit.}

\label{fig:NHdistribution}
\end{figure}
The distribution of intrinsic absorption
(Fig.~\ref{fig:NHdistribution}) shows that the Seyfert
galaxies in the \integral sample are evenly distributed between
$10^{21} \rm \, cm^{-2}$ and $3 \times 10^{23} \rm \, cm^{-2}$, as
already seen in the first \integral  AGN catalogue
(\cite{2006ESASP.604..777B}) and in the {\it Swift}/BAT AGN survey (\cite{Tueller08}). 
A significant amount of sources have reported $N_{\rm H}$ values of $N_{\rm H} <
10^{21} \rm \, cm^{-2}$. We assigned $N_{\rm H} = 10^{21}
\rm \, cm^{-2}$ to these objects, which causes the peak in the distribution.
The claim of two distinct groups resulting in two peaks in the $N_{\rm H}$ distribution as reported by Paltani et al. (2008) appears therefore caused by low number statistics. 
Although a clear dependency is seen on the intrinsic absorption when
dividing the sample in two groups, there is no significant linear correlation with any of
the parameters we tested. We summarize the analysis of the
correlations in Table~\ref{correlations}. Here the correlation
 probability is given if larger than $90 \%$. The correlations
 between Eddington ratio and $L_X$ and $M_{BH}$ are intrinsic,
 i.e. caused by the dependence of $\lambda$ on these two parameters.

For synthesis studies of the
cosmic X-ray background one will have to take into account that both,
the fraction of Compton thick and of absorbed sources, appears to be
lower than assumed in most models published so far (e.g. Treister \& Urry 2005, 
Gilli et al. 2007). Although the average redshift in the
\integral sample has increased significantly since the first catalogue
from $\langle z \rangle = 0.01$ to now $\langle z \rangle = 0.03$, \integral
will not be able to probe evolutionary effects. The only possibility
of measuring evolutionary effects would be the combination of {\it
 Swift}/BAT data with a deep ($\sim 10 \rm \, Msec$) \integral
IBIS/ISGRI field (Paltani et al. 2008).  

A systematic analysis of all {\it BeppoSAX} observations of AGN
 has been presented by Dadina (2007). Thirty-nine objects from this work are
 in common with the \integral sample. Four sources
 show strong differences in flux by more than a factor of 2
 between
 the measurements, i.e. the Seyfert~2 galaxies
 NGC~4388, NGC~2110, NGC~7172, and the type 1 radio galaxy
 3C~111. All these objects are known to be variable at hardest X-rays
 (e.g. Beckmann et al. 2007). In addition, the average flux in the
 20--100 keV band appears to be higher in the \integral data than in
 the {\it BeppoSAX} one. This can be caused by the different time
 coverage: while {\it BeppoSAX} observations are mostly snapshots of
 the sources with duration of $\Delta t \ll 100 \rm \, ks$, the {\it
 INTEGRAL} observations of the AGN have the average exposure of $1200 \rm \, ks$. Thus
 the data presented here can include some short-term bright states,
 which would have been missed in most cases by {\it BeppoSAX}. In
 addition, the fluxes presented here are model fluxes, and because more
 complex modelling is necessary for the
 {\it BeppoSAX} spectra, applying a cut-off and a reflection
 component in most cases, this leads to lower broad-band model fluxes
 compared to the simple power law model applied here (see also
 Section~\ref{NHcorrelations}). In addition, there is a systematic difference in the calibration of
the ISGRI and PDS detectors, with steeper and higher-normalization Crab spectra extracted for ISGRI (\cite{Kirsch05}). This explains also larger spectral slope values fitted to the ISGRI spectra, when compared with the values quoted by Dadina (2007), although the primary reason for
the these differences is the simpler spectral model applied to the
 \integral data.

{\it CGRO}/OSSE (\cite{OSSE}) covered the energy range of
approximately $50-10^3$ ~keV. It therefore primariliy detected AGN with hard and bright
X-ray spectra, which we also expect to be detectable by {\it
 INTEGRAL}. While the first {\integral} AGN catalogue listed 24
OSSE-detected AGN not seen by {\it INTEGRAL,} this number has now
decreased to 10. We list those sources and the IBIS/ISGRI exposure time on the particular AGN in
Table~\ref{OSSEnonINTEGRAL}. It can be expected that the persistent
sources in this list, i.e. the Seyfert and starburst galaxies, will be
detected once a significant amount of exposure time is available for
these objects. 

As expected, only a few objects are jointly detected by \integral
IBIS/ISGRI and {\it Fermi}/LAT, according to the bright source list
based on three months of LAT data (\cite{FermiLAT}). These 13 objects
are blazars, except for the two radio galaxies Cen~A and
NGC~1275. The common blazars between ISGRI and LAT are 1ES~0033+595, PKS~0528+134, QSO~B0716+714, Mrk~421,
3C~273, 3C~279, Mrk~501, PKS~1830-211, 1ES~1959+650, BL~Lac, and
3C~454.3. The group of sources jointly detected by {\it Fermi}/LAT
and ISGRI is expected to increase significantly through \integral
target-of-opportunity (ToO) observations of blazars that show a flare
in the LAT data. 

\subsection{The intrinsic hard X-ray spectrum}
\label{NHcorrelations}

The effect that Seyfert 1 and low-absorbed objects appear to have steeper
X-ray spectra than the Seyfert 2 and highly absorbed AGN was first
noticed by Zdziarski et
 al. (1995), based on {\it Ginga} and {\it CGRO}/OSSE data and later
 confirmed e.g. by Gondek et al. (1996) using combined {\it EXOSAT},
 {\it Ginga}, {\it HEAO-1}, and {\it CGRO}/OSSE spectra. 
A study of {\it BeppoSAX} PDS spectra of 45 Seyfert galaxies came to
a similar conclusion, although the spectra of Seyfert 2 appeared
steeper when considering a possible cut-off in the spectra of
Seyfert~1 galaxies 
(\cite{Deluit03}). 
The difference in the hard X-ray spectral slope between Seyfert 1 and 2
 has been a point of discussion ever since its discovery. Zdziarski et
al. (2000) considered the anisotropy of Compton scattering in planar
geometry and effects of reflection, but came to the conclusion that
this cannot be the sole explanation. 
Beckmann et al. (2006) argued that the difference might be a selection
effect, as objects have to have a harder X-ray spectrum to
be detectable when strong intrinsic absorption is present. With the growing sample of AGN, this argument does not
seem to hold, as the absorption in the energy band $>20 \rm \, keV$ is
negligible for Compton thin objects, and also the ongoing
identification effort of newly detected hard X-ray sources did not
reveal a different population than already presented in Beckmann et
al. (2006).

A solution might be provided when considering the effects of Compton
reflection on the hard X-ray spectrum, as shown in the
  previous section. Recent analysis of a sample of 105 Seyfert galaxies using the spectra
collected
with {\it BeppoSAX} in the 2--200 keV band (Dadina 2008) provided no evidence
of any spectral slope difference when applying more complex model
fitting including a reflection component (PEXRAV). The mean photon index values
found for Seyfert 1 and Seyfert 2 samples were $\Gamma = 1.89 \pm 0.03$ and
$\Gamma = 1.80 \pm 0.05$. The difference between types 1 and 2 is seen in
this model in the different strength of the reflection component, with
$R = 1.2 \pm 0.1$ and $R = 0.9 \pm 0.1$, and different cut-off
energies of $E_C = 230 \pm 22 \rm \, keV$ and $E_C = 376 \pm 42 \rm \,
keV$, for Seyfert~1 and Seyfert~2, respectively. It has to be
pointed out that spectral slope, reflection strength, and cut-off
energy are closely linked.   
The IBIS/ISGRI data have a
  disadvantage over broad-band data when studying 
  the spectral shape of the hard X-ray continuum, as we
  lack information about the spectrum below 18 keV. But on the
  positive side, using data from only one instrument, the spectra do not suffer
  from the problem of intercalibration factors, which is apparent in
  all studies using different instruments, and especially when using
  data from different epochs. In those cases, where spectra are taken
  by more than one instrument at different times, the flux variability can
  mimic a stronger or weaker reflection component or cut-off energy
  (e.g. Panessa et al. 2008).

The \integral data show consistent slopes for the spectra
of unabsorbed / type 1 and absorbed / type 2 objects already
when a simple power-law model is used.
When applying the model used by Dadina (2007) to the stacked
 \integral spectrum of Seyfert galaxies, we get similar results:
 the underlying powerlaw appears to have consistent (within $2\sigma$) spectral slope
 for type 1 ($\Gamma = 1.96$) and type 2 ($\Gamma = 1.91$) objects
 and the same reflection strength $R \simeq 1.1$.
The data do not allow to determine the cut-off energy or
inclination angle when fitting the reflection component
(Fig.~\ref{fig:stackSy}). When fitting a simple cut-off power law, the
\integral data show the same trend as the {\it BeppoSAX} sample,
i.e. a lower cut-off energy for Seyfert 1 ($E_C = 86 \rm \, keV$) than
for Seyfert 2 ($E_C = 184 \rm \, keV$). It has to be taken
  into account, though, that the fit to the Seyfert~2 data is bad
  quality, and that fixing the cut-off here to the same value as
  derived for the Seyfert~1, also leads to the same spectral slope. 
When fitting a reflection model to the stacked data, one gets a consistent
photon index of $\Gamma \simeq 1.95$ and reflection strength $R
\simeq 1.3$
for both absorbed and unabsorbed AGN.
These values of $\Gamma$ and $R$ agree with the
correlation $R = (4.54 \pm 1.15) \times \Gamma - (7.41 \pm 4.51)$
Dadina (2008) found for the {\it BeppoSAX} AGN sample, which, for the
\integral sample with $\Gamma = 1.95$, would lead to $R = 1.4$. This
$R(\Gamma)$ correlation was first noted based on {\it Ginga} data for
extragalactic and Galactic black holes,
leading to $R = (1.4 \pm 1.2) \times 10^{-4} \, \Gamma^{(12.4 \pm 1.2)}$
(\cite{Zdziarski99}), which in our case would result in a smaller
expected reflection component with $R = 0.6$ but within $1 \sigma$ of
the value detected here. Absorbed and unabsorbed show a consistent
turnover at about $E_C = 100 \rm \, keV$ when a cut-off power law
model is applied.

The observed dichotomy of different spectral slopes for
type 1 and type 2 objects might therefore be caused by data with too
low significance, which do not allow to fit the reflection component, 
or in general by a strong dependence of the spectral slope on the choice of the
fitted model.
One aspect that has to be kept in mind is the dependence
of the reflection strength $R$ on the model applied and on the
geometry assumed. 
Murphy \& Yaqoob (2009) showed recently that their model of a reflection spectrum from
a Compton-thick face-on torus that subtends a solid angle of $2\pi$
at the X-ray source is a factor of $\sim 6$ weaker than that expected
from a Compton-thick, face-on disc as modelled in PEXRAV. Therefore,
applying a torus model to the data presented here would result in much
less reflection
strength. 

\subsection{Eddington ratios and accretion rates in Seyfert galaxies}
A different approach to search for differences or similarities in
Seyfert galaxies is to study the accretion rates.
Middleton et~al. (2008) suggest that different accretion states
lead to differences 
in the hardness of hard X-ray ($E > 10 \rm \, keV$) spectra 
between types 1 and 2 AGN (type~1 spectra
being systematically softer).
Using data from {\it CGRO}/OSSE, {\it
  BeppoSAX}, and \integral, they found that the 24 Seyfert~2
galaxies in their sample of hard X-ray selected AGN show an accretion
rate (parameterized with the Eddington ratio) in average smaller than that
of their 23 Seyfert~1. This effect has also been seen when
studying AGN detected by {\it Swift}/BAT (\cite{Winter09}).
This would be consistent with all accreting black holes in general showing
harder spectra at low accretion rates (\cite{Laor00}, \cite{Remillard06}).

\begin{figure}
\includegraphics[height=8.5cm,angle=90]{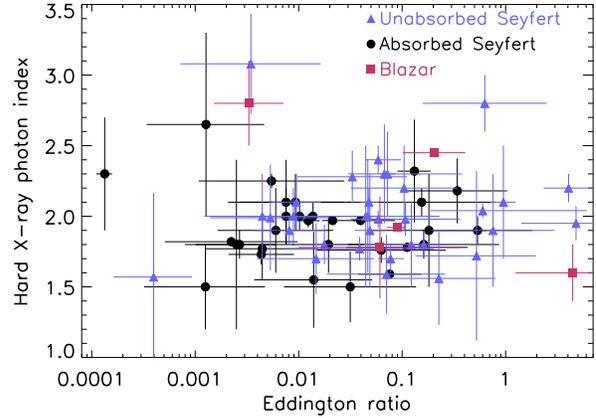}
\caption[]{Photon index of a simple power-law fit to the IBIS/ISGRI
 data versus Eddington ratio $\lambda$.}

\label{fig:Eddington_gamma}
\end{figure}

\begin{figure}
\includegraphics[height=8.5cm,angle=0]{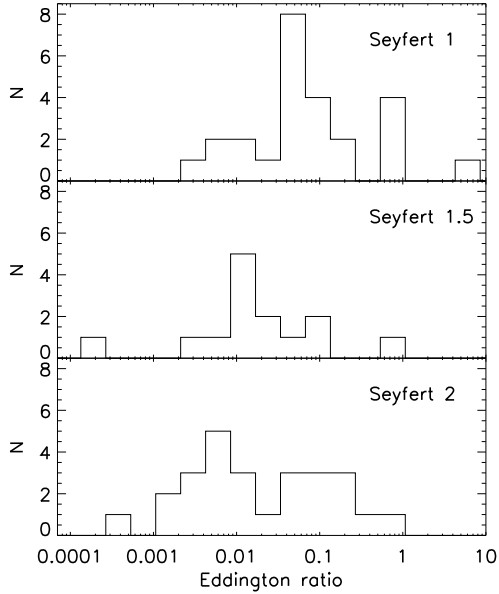}
\caption[]{Distribution of Eddington ratios for Seyfert~1, 1.5, and
 type 2 AGN
 in the sample.}

\label{fig:Eddington_distribution}
\end{figure}

The $20 - 100 \rm \, keV$ luminosities as derived from the IBIS/ISGRI data were used to approximate the bolometric luminosity. Assuming
a canonical photon index of 2.0 for a single power law, the total X-ray
luminosity  is about $L_{(1-200 \rm \, keV)} = 3 \times L _{(20-100
 \rm \, keV)}$. Assuming that the first peak of the spectral energy
distribution 
is as strong as the X-ray luminosity, we derive $L_{Bol} = 6 \times L
_{(20-100 \rm \, keV)}$. The Eddington luminosity considering pure
hydrogen is given by $L_{Edd}
= 1.26 \times 10^{38} \frac{M}{M_\odot} \rm \, erg \, s^{-1}$, and thus
the Eddington ratio is $\lambda = L_{Bol} / L_{Edd} = 4.8 \times
10^{-38} \, L_{(20-100 \rm \, keV)} \, \frac{M_\odot}{M} \rm \,
erg^{-1} \, s$. 
The Eddington ratio can be computed for 71 Seyfert
objects of the sample presented here. The photon
index as determined from IBIS/ISGRI data and the Eddington ratio are
compared in Fig.~\ref{fig:Eddington_gamma}. The four  AGN on the top left of this figure, with steep spectra and low
  Eddington ratio ($\lambda < 0.005$), are the Seyfert~1 1H~1934-063,
  the Seyfert~2 objects SWIFT~J0601.9-8636 and NGC~4258, and the
  blazar Mrk~501. There is no significant
correlation detectable, even when excluding these outliers. 
As observed by Steffen et al. (2003) in {\it Chandra} data of AGN, it
appears that the 2--8 keV luminosity function is dominated by type 1
AGN at high X-ray luminosities and by type 2 at low luminosities. The
same effect is seen also in the \integral luminosity function
(\cite{XLF}). Connected to this, we observe not only that the absorbed
sources are less luminous than the unabsorbed ones, but also that
absorbed sources have smaller accretion rates as seen in lower Eddington ratios (Fig.~\ref{fig:Eddington_distribution}).
In agreement with Middleton et~al. (2008) results, 
we find that the average values of Eddington ratio for
Seyfert~1 ($\langle \lambda_{\rm Sy1} \rangle = 0.064$) are higher than those found for intermediate Seyfert
type ($\langle \lambda_{\rm Sy1.5} \rangle = 0.015$) and
Seyfert~2  with $\langle \lambda_{\rm Sy2} \rangle = 0.02$ (Fig.~\ref{fig:Eddington_distribution}), although we
  do not observe the differences in the underlying spectra, as seen in 
  their study. The same applies for
the separation into unabsorbed ($\langle \lambda_{(N_{\rm H} < 10^{22} \rm
 cm^{-2}) } \rangle = 0.06$) and absorbed sources ($\langle \lambda_{(N_{\rm H} > 10^{22} \rm
 cm^{-2}) } \rangle = 0.015$). 
To calculate the probability that the Eddington ratios of Sey1/unabsorbed objects
and Sey2/absorbed AGN are drawn from the same population, we applied a
Kolmogorov-Smirnov test. We can reject the null hypothesis (same population), with a probability
of false rejection of 0.1\% and 3\% for the Seyfert 1 --  Seyfert 2
and unabsorbed -- absorbed objects, respectively.
On the other hand, since we do not find a significant correlation between the hard X-ray photon index
and the Eddington ratio, 
we cannot back up the scenario by Middleton et al. (2008) as a way to explain
the different spectral hardness of type 1 and type 2 AGN.

\begin{figure}
\includegraphics[height=8.5cm,angle=0]{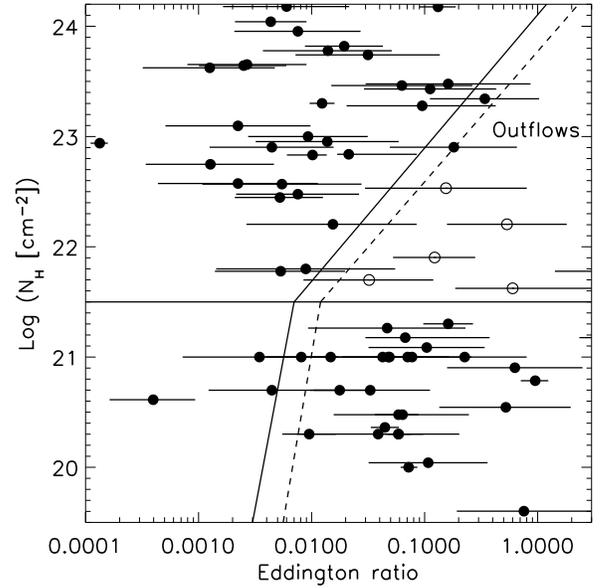}
\caption[]{Intrinsic absorption versus Eddington ratio for the Seyfert
 galaxies in the \integral sample. The 'forbidden zone' described by Fabian et al. (2009) is
 occupied by 5 {\integral} AGN (empty circles).}

\label{fig:Nh_Edd_Fabian}
\end{figure}

When studying the effect of radiation pressure on dusty absorbing gas
around AGN, Fabian et al. (2008) trace a region in the $N_{\rm H}$ -
Eddington ratio plane that is forbidden to long-lived clouds in AGN.
In fact, even when the AGN is in the sub-Eddington regime for the ionized gas,
it can appear to be super-Eddington for the dusty gas, hence
ejecting the surrounding absorbing clouds.
Objects in this region of the $N_{\rm H}$ - Eddington ratio plane could
present outflows or show transient or variable absorption.
Fabian et al. (2009) tested the predictions of this model using the
\textit{Swift}/BAT AGN sample of Winter et al. (2008) and find a good
agreement, with only 1 object lying in the forbidden
area, the Seyfert 1.9 object MCG--05--23--016. This object has been
shown to exhibit high accretion rates at the Eddington limit (Beckmann et al. 2008).
In Fig.~\ref{fig:Nh_Edd_Fabian} we show the regions defined by Fabian et al. (2009) in the
$N_{\rm H}$ - Eddington ratio plane and our AGN sample. 
The forbidden
region is on the upper-righthand side of the plot and bordered by an upper
limit to $N_{\rm H}$ (for absorption due to dust lanes at
few kpc from the AGN centre) and the Eddington limit for the dusty gas when only the black hole mass is considered to be gravitationally
important (continuous line) or if also as much mass from intervening
stars is included (dashed line).
It is important to keep in mind that the bolometric luminosity is a
critical parameter in defining the limits of the forbidden area, as
different estimates of it can be used.
Therefore, using a bolometric correction by a factor of few larger than
what we applied here would shift the data points towards higher Eddington ratios,
occupying the forbidden region completely.
Nevertheless, with the bolometric luminosity we estimate ($L_{Bol} = 6
\times L_{(20 - 100 \rm \, keV)}$), and find 
5 objects with a detection significance
   $> 4 \sigma$ in this area: IGR~J00335$+$6126, MCG--05--23--016,
Mrk~766, IC~4329A, and NGC~5506. Out of these, only
   MCG--05--23--016 occupies the forbidden region also in Fabian et
   al. (2009), whereas IGR~J00335$+$6126 was not included and
    Mrk~766 lacked absorption information in their study, IC~4329A is
   located close to the border of the
   forbidden zone, and NGC~5506 is not within this area. This
   raises the question of whether this region indeed cannot be occupied
   persistently by Seyfert type AGN, although the 5 objects do display
   some peculiarities, as described in the following.
IC~4329A and IGR~J00335$+$6126 are just above the upper limit $N_{\rm H}$
and MCG--05--23--016 has a complex spectrum with warm absorbers, not
necessarily related to dusty gas (Fabian et al. 2009). 
The Seyfert~1.9 NGC~5506 has been identified as an
obscured narrow-line Seyfert 1 (\cite{Nagar02}), which might explain the high accretion
rate in this object. And the Seyfert~1.5 Mrk~766 with a black hole mass of
only $3.5 \times 10^6 \rm \, M_\odot$ (\cite{Uttley05}) might be a similar case to the
highly efficient Seyfert MCG--05--23--016
(\cite{Beckmann08}). 
But the small number of objects and the fact that the uncertainty on
the Eddington ratio is still rather large do not allow us
to conclusively state that strongly absorbed Seyfert galaxies cannot
exhibit high accretion rates over a long phase of their lifetimes.

\subsection{Correlations with optical data and \aox}
\label{opticalcorrelations}

\begin{figure}
\includegraphics[height=8.5cm,angle=90]{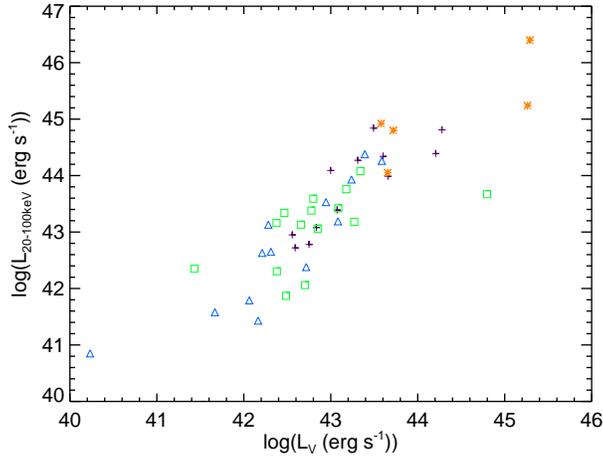}
\caption{X-ray luminosity \lx\ versus V-band luminosity \lv. AGN of different types are
 located along the same dispersion line, over more than 5 decades in
 luminosity. Symbols: crosses - Seyfert 1-1.2; triangles - Seyfert
 1.5; squares - Seyfert 1.8-2; asterisks - BL Lacs.}
\label{fig:Lx_Lv}
\end{figure}

We show in Fig.~\ref{fig:Lx_Lv} the dispersion diagram of \lx\ vs. \lv, with
different symbols for the different classes of objects considered. While the
apparent correlation between both luminosities is certainly driven by the
distance effect evident in Fig.~\ref{fig:z_Lx}, it is remarkable that the
different classes of objects, from Seyfert 2 to blazars, are located on
the same correlation line, over more than 5 decades in
luminosity. Subtracting the common dependence on redshift 
through a partial correlation analysis, the correlation between the luminosities is still
statistically significant, with a correlation coefficient of 0.74 and the
probability of a chance occurrence is $\ll 0.01\%$. To test this
correlation further,
we made simulations using a bootstrap method (\cite{Simpson86}): to each couple
of X-ray flux and redshift we randomly assigned an optical flux, drawn from the real values
found for our sample and without excluding multiple choices of the same value
(see also \cite{Bianchi09} for a similar procedure). 
We then computed the X-ray and optical luminosities
and calculated the Spearman and the partial correlation coefficients for the \lx\ -- \lv\
relation and the Spearman coefficient for the $F_X - F_V$ relation. Repeating
this procedure 100,000 times, we were able to build histograms of the correlation coefficients 
of the simulated samples and found that only 0.001\% of the simulated samples have a 
(Spearman or partial) correlation coefficient greater than measured in the real sample for the \lx\ -- \lv\
relation. 
For the $F_X - F_V$ correlation, the probability of chance occurrence is higher but 
still not significant, $\sim 2\%$. This indicates that the 
X-ray and optical emissions are indeed correlated, beyond the bias introduced by the common
dependence on distance.

\begin{figure}
\includegraphics[height=8.5cm,angle=0]{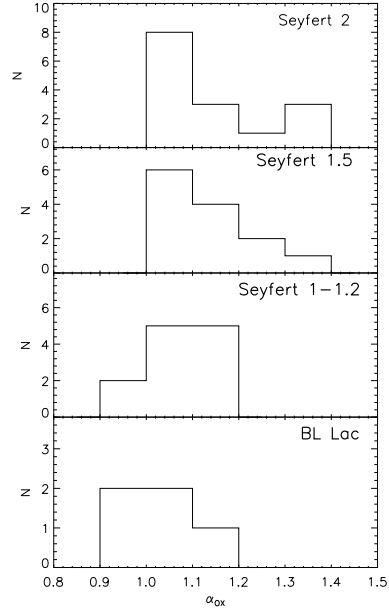}
\caption{ \aox\ histograms for 4 subtypes of AGN.}
\label{fig:aox_histogram}
\end{figure}

To further investigate this relation,
we computed the histograms of the \aox\ values for the different subtypes.
We show in Fig.~\ref{fig:aox_histogram} that the peaks of the distributions
coincide for the four subsamples, with mean values close to each other
(Seyfert~2: 1.14, Seyfert~1.5: 1.13, Seyfert~1-1.2: 1.08, and BL~Lac: 1.03). Seyfert 2
and 1.5 nevertheless show an extended wing towards higher \aox\ ratios.
NGC~1068, the prototypical Seyfert 2 galaxy, shows the highest \aox\ value,
1.38. Because it is a Compton-thick object, this high \aox\ value can
be understood clearly. Other Compton-thick Seyfert galaxies, such as NGC~3165 and ESO138--1,
also show \aox\ values above 1.20. But other objects with high \aox\ values,
such as NGC~1052 and NGC~5033, have low hydrogen column densities; 
in contrast several objects with high neutral columns, like NGC~3281, NGC~4945, and
ESO~383--18, show relatively low \aox\ values. 

As discussed in Sect.~\ref{opticalvariability}, only 3
objects in the sample have shown significant optical variability over the period being monitored
($\Delta V > 0.5$ mag).
For the blazars QSO~B0716+714 and 3C~279, both their \lv\ and \lx\ values have 
to be dominated by the central AGN, with only a minor contribution by stars in V. 
This is indeed expected for this type of high-luminosity blazars
while accretion processes and bulge stars 
dominate the V band in less active Seyfert galaxies.

That most Seyfert galaxies in the sample look similar when
analysed from the point of view of their \aox\ value indicates that the true
nature of these objects is indeed very similar, thus supporting the unified
scenario of AGN. The central black hole behaves very similarly with respect to
the host galaxy, independent of the type of object. The classification as type
1 or type 2 AGN would mainly derive from observations of parameters
dependent on geometrical effects, such as the profile of the emission lines
or of the X-ray emission,
but would not be tracing systematic
differences in the intrinsic nature of these objects. 

\subsection{A fundamental plane of AGN activity}
\label{fundamentalplane}

In the view of unification of different AGN types, it has been pointed
out that  AGN, spanning black hole masses in the range of
$10^5 M_\odot \lae M_{\rm BH} \lae 10^9 M_\odot$ and even accreting black holes in
X-ray binaries with $M_{\rm BH} \sim 10 M_\odot$, show similarities in
radiative efficiency and jet power versus accretion rate (e.g. Fender
et al. 2007). This connection gave rise to the `fundamental plane' of
black hole activity. It has been found that indeed there is a close
connection between the radio and X-ray luminosity of Galactic and
super massive black holes of the form $L_{\rm radio} \propto
L_X^{0.7}$ (Corbel et al. 2000, 2003), linking the jet activity to the
total output of the central engine. Later on, a connection of these
two parameters with the black hole mass itself was found, establishing
the fundamental plane of AGN and Galactic black hole activity in the
form of $L_{\rm radio} \propto L_X^{0.6} M_{\rm BH}^{0.8}$ (e.g. \cite{Merloni03}). 

\begin{figure}
\includegraphics[height=8.5cm,angle=90]{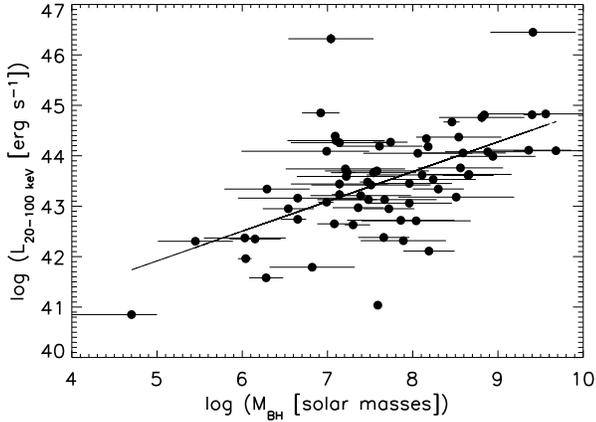}
\caption[]{Hard X-ray luminosity versus black hole masses for Seyfert
  galaxies. Errors on $L_X$ are smaller than the symbols. More massive AGN appear to be brighter in the hard
  X-rays with $L_X \sim M_{BH}^{ \,\, 0.7 \pm 0.1}$. The correlation coefficient is 0.52 with a probability of
  non-correlation $< 2 \times 10^{-6}$.}

\label{fig:BHmass_Lx}
\end{figure}

The correlation between X-ray luminosity and black hole mass is
  also common to all objects in our sample, showing
  indeed that more massive AGN are more
luminous 
($L_X \sim M_{BH}^{ \,\, 0.7 \pm 0.1}$, as shown in Fig.~\ref{fig:BHmass_Lx}).
A similar relation has lately been reported for various X-ray selected
AGN samples (e.g. \cite{Bianchi09}, \cite{Wang09}), and also
in these cases the slope is lower than 1. This could indicate that
more massive black holes either have
lower accretion rates than less massive objects, or a smaller fraction of their total power is 
converted into X-ray luminosity (assuming that the bolometric luminosity scales linearly with black hole mass). 
As reported in Table~\ref{correlations}, a significant correlation is also detected between the black hole mass
   and the optical luminosity. Only for 8
out of 41 objects might the correlation of $L_V$ versus $M_{BH}$ be
induced by the method used to estimate the mass, i.e. when deriving
the mass from the K-band magnitudes (assumed to represent the bulge
luminosity) or from continuum optical luminosity. 
However, caution should be used in general when considering a correlation between luminosity and black hole mass,
because possible selection effects or different biases could contribute to the observed correlation in non complete samples.
Woo \& Urry (2002) argue that, when correlating bolometric luminosity with black hole mass, the Eddington luminosity
sets a (soft) upper limit to the luminosities and therefore determines the empty region of the diagram in the upper left
corner. On the other hand, the lower-right corner should be populated by low-luminosity, massive black holes that are
not included or are rare in the high-energy AGN samples, such as normal or radio galaxies.

As discussed in detail in the previous section, there is also a
  significant correlation between optical and X-ray luminosity. Therefore, as summarised in
Table~\ref{correlations}, we found three significant correlations
 in our sample between the luminosities $L_X$ and $L_V$, and the mass
 of the central black hole $M_{\rm BH}$. This leads to the assumption
 that also these parameters, similar to $L_X$, $L_R$, and $M_{\rm
  BH}$, form a fundamental plane for AGN. By applying an analysis
  following Merloni et
 al. (2003), we fit the
 data with the function
\begin{equation}
\log L_V = \zeta_{VX} \, \log L_X + \zeta_{VM} \, \log M_{\rm BH} +
b_V  .
\end{equation}
We obtain $\zeta_{VX} = 0.59 \pm 0.07$, $\zeta_{VM} = 0.22 \pm 0.08$, and $b_V = 16.0$,
leading to 
\begin{equation}
\log L_V = 0.59 \log L_X + 0.22 \log M_{BH} + 16.0;
\end{equation}
i.e., $L_V \propto L_X^{0.6} M_{BH}^{0.2}$.  
\begin{figure}
\includegraphics[height=8.5cm,angle=90]{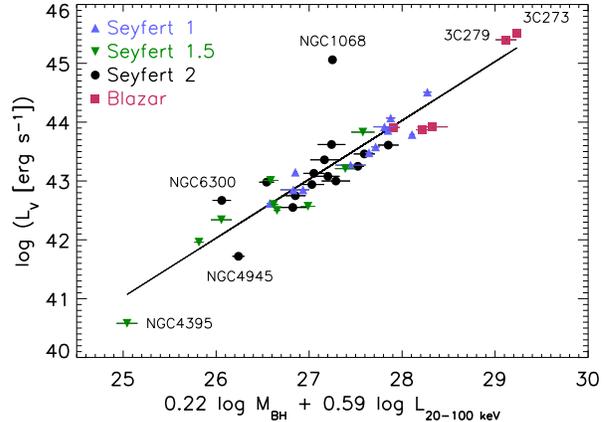}
\caption[]{Fundamental plane of optical luminosity $L_V$, X-ray
 luminosity $L_X$, and mass of the central black hole $M_{BH}$. Errors on luminosity are smaller than the symbols.}

\label{fig:fundamentalplane}
\end{figure}
We show this relation in Fig.~\ref{fig:fundamentalplane}. 
As already pointed out and discussed here and in Sect~\ref{opticalcorrelations},
we carefully investigated that the effects of distance and selection 
effect are not causing the correlation observed here.

While the fundamental plane including the radio and X-ray luminosity
can be understood as a connection of the jet activity, as visible in
the radio, and the
accretion flow, as dominating the X-rays, the correlation found here
shows a different connection. The optical luminosity is commonly
  thought to be dominated by the AGN accretion disc (e.g. \cite{Siemiginowska95}) and therefore by the
accretion processes onto the supermassive black hole, but there is
also a possible contribution by the jet (\cite{Soldi08}) and emission of the bulge, and therefore of the stars in the host galaxy
contributing to it. The latter is especially important as the resolution of
\integral's OMC camera does not allow a deconvolution of the core and
the bulge. Nevertheless it shows that there is a significant
bulge-$M_{BH}$ correlation and that accretion processes are closely
linked to the mass of the central black hole. The finding that
  this fundamental plane holds for all Seyfert types indicates
  further that these AGN are indeed intrinsically the same.

\section{Conclusions}
\label{conclusions}

We have presented the second \integral AGN catalogue, including 187
extragalactic objects. The AGN population detected by \integral is
dominated by Seyfert galaxies in the local ($\langle z \rangle =
0.03$) universe, with moderate X-ray luminosity ($\langle L_{20 - 100
 \rm keV} \rangle = 4 \times 10^{43} \rm \, erg \, s^{-1}$). Seyfert~1 galaxies appear to
have 
higher luminosities ($\langle L_{20 - 100
 \rm keV} \rangle = 10^{44} \rm \, erg \, s^{-1}$) and Eddington ratio
($\langle \lambda_{\rm Sy1} \rangle = 0.064$) than the Seyfert~2 galaxies 
($\langle L_{20 - 100 \rm keV} \rangle = 2.5 \times 10^{43} \rm \, erg \, s^{-1}$,
$\langle \lambda_{\rm Sy2} \rangle = 0.02$). Although IBIS/ISGRI
  spectra alone lack the information about the iron line
  complex and the continuum shape below 18 keV, they can be used to
  study the average Seyfert spectra in a statistical way.
The underlying continuum of the hard X-ray spectrum appears to be
 consistent
between different Seyfert types, both when a simple power-law model
is applied and 
 when considering the effects of Compton
 reflection. Applying the PEXRAV reflection model with no high energy
 cut-off, the Seyfert~1 and 2 galaxies show the same underlying power
 law with $\Gamma \simeq 1.95$ and a reflection component of $R \simeq
 1.1$, when applying different inclination angles of $i \simeq 30^\circ$ and
 $i \simeq 60^\circ$, respectively.
Although, when applying a cut-off power law model to the stacked spectra, the Seyfert~1 show lower
cut-off energies ($E_C = 86 {+21 \atop -14} \rm \, keV$) than the
Seyfert~2 objects ($E_C = 184 {+16 \atop -52} \rm \, keV$), the bad
quality of the fit in the latter case and that fixing the
cut-off to the value of the Seyfert~1 leads to a similar spectral
slope might indicate that the spectra are intrinsically indeed the same.

The same differences as for different Seyfert classes are observable when considering the intrinsic absorption:
the unabsorbed sources also have 
higher luminosities ($\langle L_{20 - 100 \rm keV}\rangle  = 6.3
\times 10^{43} \rm \, erg \, s^{-1}$) and Eddington ratio
($\langle \lambda \rangle = 0.06$) than the absorbed AGN 
($\langle L_{20 - 100 \rm keV} \rangle = 2.5
\times 10^{43} \rm \, erg \, s^{-1}$, $\langle \lambda \rangle =
0.015$).
Also separating the objects into absorption classes, the underlying
continuum appears similar when considering the effects of Compton reflection.
The mass of the
central black hole is on average the same among the different Seyfert
types and absorption classes, with $\langle M_{BH} \rangle = 4 \times 10^7 \rm
\, M_\odot$. Also comparing optical to hard X-ray emission, the different
Seyfert classes show the same ratio (\aox = 1.1). On average, the hard X-ray spectra of Seyfert~1.5 objects
are closer to those of the Seyfert~1 class than to Seyfert~2.

The optical data provided by {\it INTEGRAL}/OMC can be used to monitor
variability. Strong variability ($\Delta V \gae 0.5 \rm \, mag$) is
only seen in three objects within the optical sample of 57 AGN, i.e. in the blazars QSO
B0716+714 and 3C 279, and in NGC~4151.


The overall picture can be interpreted within the scenario of a
unified model.
The whole hard X-ray detected Seyfert population fills the parameter
space of spectral shape, luminosity, and accretion rate smoothly, and
only an overall tendency is seen in which more massive objects are
more luminous, less absorbed, and accreting at higher Eddington ratio. An
explanation for why the absorbed sources have been claimed to
show flatter spectra in the hard X-ray domain when fit by a simple power law 
can be 
that the slope of the continuum strongly depends on the fitted model and
that Compton reflection processes play a major role here. Considering these
effects, it appears that the different Seyfert types are indeed
intrinsically the same. 

More evidence for the unified scheme is that a fundamental plane can be found between the mass of the central
object and optical and X-ray luminosity. The correlation takes
  the form $L_V \propto L_X^{0.6} M_{BH}^{0.2}$,
 similar to what is found in previous studies between $L_R$, $L_X$, and $M_{BH}$. This links the
accretion mechanism with the bulge of the host galaxy and with the
mass of the central engine in the same way in all types of Seyfert
galaxies. The connection is also apparent through the same optical-to-hard
X-ray ratio measured in all Seyfert classes.

Evolutionary effects are likely to be beyond the AGN population
accessible by \integral and {\it Swift}. Deep hard X-ray surveys by
future missions like {\it NuSTAR}, {\it Astro-H}, and {\it EXIST}
will be able to answer this question through deep observations of
small portions of the sky. 

\begin{acknowledgements}
{\it INTEGRAL} is an ESA project funded
by ESA member states (especially the PI countries: Denmark, France,
Germany, Italy, Spain, Switzerland), Czech Republic, Poland, and with the
participation of Russia and the USA. We thank the anonymous referee
for the comments that helped to improve the paper. 
This research has made use of data obtained through the High Energy
Astrophysics Science Archive Research Center Online Service, provided
by the NASA/Goddard Space Flight Center. We acknowledge the use of
public data from the {\it Swift} data archive and from the {\it
  INTEGRAL} data archive provided by the ISDC. SS acknowledges the
support by the Centre National d'Etudes Spatiales (CNES).
PL and AAZ have been supported in part by the Polish MNiSW
grants NN203065933 and 362/1/N-INTEGRAL/2008/09/0, and the Polish
Astroparticle Network 621/E-78/BWSN-0068/2008.
JMMH, AD and JA
are supported by the Spanish MICINN grant ESP2008-03467.
\end{acknowledgements}

\onecolumn
\begin{longtable}{lccrrrrcc}
\caption{{\integral} AGN catalogue\label{catalog}. 
 The column {\it
   Method} indicates the method used to determine the black hole
 mass, see Sect.~\ref{bhmasses} for details.
}\\
\hline\hline
Name &  Type & $z$ &  R.A. &  DEC & ISGRI &  JEM-X & $\log M_{BH}$ & Method\\
 &   &  & $[\rm deg]$ & $[\rm deg]$ & $[\rm ksec]$ &  $[\rm ksec]$ & $[M_\odot]$\\ 
\hline
\endfirsthead
\caption{continued.}\\
\hline\hline
Name &  Type & $z$ &  R.A. &  DEC & ISGRI &  JEM-X & $\log M_{BH}$ & Method\\
 &   &  & $[\rm deg]$ & $[\rm deg]$ & $[\rm ksec]$ &  $[\rm ksec]$ &
$[M_\odot]$\\ 
\hline
\endhead
\hline
\endfoot
IGR J00040+7020$^x$ & Sy2   & 0.096  &  1.00638  &  70.32125 & 1947.3 & 20.9\\
IGR J00254+6822 & Sy2   & 0.012  &  6.38092  &  68.36147 & 2534.6 & 178.9\\
IGR J00335+6126 & Sy1   & 0.105  &  8.3265   &  61.46178 & 3338.7 &
356.5 & $8.5 \pm 0.5^a$ & LL/CL\\
1ES 0033+59.5   & BLLac & 0.086  &  8.96929  &  59.83461 & 3338.7 & 356.4\\
Mrk 348         & Sy2   & 0.0151 & 12.19642  &  31.95697 &  150.7 &
21.1 & $7.2 \pm 0.7^b$ & SO\\
NGC 418$^x$         & AGN   & 0.0190 & 17.64842  &--30.22128 &   38.5 &  --\\
NGC 526A        & Sy1.5 & 0.0191 & 20.97583  &--35.06528 &   73.7 &
-- & $8.1 \pm 0.7^c$ & S\\
ESO 297-18      & Sy2   & 0.0252 & 24.65492  &--40.01131 &   59.5 &
-- & $9.7 \pm 0.5^n$ & KM\\
IGR J01528-0326 & Sy2   & 0.0167 & 28.20375  & --3.44749 &  837.1 &  79.6\\
NGC 788         & Sy2   & 0.0136 & 30.27687  & --6.81553 &  926.6 &
117.9 & $7.5 \pm 0.7^b$ & SO\\
Mrk 590         & Sy1.2 & 0.0264 & 33.63984  & --0.76669 &  936.0 &
266.6 & $7.14 {+0.1 \atop -0.09}^d$ & R\\
IGR J02097+5222 & Sy1   & 0.0492 & 32.40700  &  52.44543 &  698.4 &   2.0\\
SWIFT J0216.3+5128 & Sy2 &0.0288 & 34.11292  &  51.42375 &  481.9 &   2.0\\
Mrk 1040        & Sy1.5 & 0.0167 & 37.06079  &  31.31094 &   42.8 &
-- & $7.6 \pm 0.3^e$ & S\\
IGR J02343+3229 & Sy2   & 0.0162 & 38.57500  &  32.48333 &   70.5 &  --\\
NGC 985         & Sy1   & 0.0431 & 38.65738  & --8.78761 &  808.1 &
48.3 & $8.9 \pm 0.5^n$ & KM\\
NGC 1052        & Sy2   & 0.0050 & 40.27000  & --8.25578 &  707.0 &
31.1 & $8.2 \pm 0.3^e$ & S\\
RBS 345         & Sy1   & 0.0690 & 40.56667  &   5.53000 &  443.2 &  --  \\
NGC 1068        & Sy2   & 0.0288 & 40.67012  & --0.01344 &  914.4 &
44.8 & $7.2 \pm 0.1^e$ & M\\
QSO B0241+62    & Sy1   & 0.0446 & 41.24042  &  62.46847 &  711.9 &  35.2\\
IGR J02466-4222 & AGN   & 0.0695 & 41.65375  &--42.36600 &  191.4 &  21.9\\
IGR J02501+5440 & Sy2   & 0.015  & 42.67417  &  54.70419 &  773.8 &  13.5\\
MCG-02-08-014   & Sy2   & 0.0168 & 43.09750  & --8.51042 &  528.1 &   4.7\\
NGC 1142        & Sy2   & 0.0288 & 43.80133  & --0.18381 &  634.1 &
-- & $9.4 \pm 0.5^n$ & KM\\
QSO B0309+411   & Sy1   & 0.136  & 48.25817  &  41.33366 &  488.4 & 303.1\\
IGR J03184-0014$^x$&QSO & --     & 49.60000  & --0.22889 &  --    &  --\\
NGC 1275        & Sy2   & 0.0176 & 49.95067  &  41.51170 &  506.8 &
282.0 & $8.5 \pm 0.7^e$ & S\\
1H 0323+342     & Sy1   & 0.0629 & 51.17150  &  34.17941 &  431.6 &   0.9\\
IGR J03334+3718 & Sy1.5 & 0.0547 & 53.32833  &  37.30305 &  463.4 &   5.6\\
NGC 1365        & Sy1.5 & 0.0055 & 53.40208  &--36.13806 &  143.3 &
3.0 & $7.7 \pm 0.3^f$ & S\\
IGR J03532-6829 & BLLac & 0.0870 & 58.30833  &--68.48306 &  740.8 &   9.4\\
3C 111          & Sy1   & 0.0485 & 64.58867  &  38.02661 &  160.1 &
-- & $9.6 \pm 0.8^c$ & B\\
3C 120          & Sy1   & 0.0330 & 68.29623  &   5.35434 &  365.9 &
79.3 & $7.7 \pm 0.2^g$ & R\\
UGC 3142        & Sy1   & 0.0217 & 70.94537  &  28.97194 &  449.6 &  39.9\\
LEDA 168563     & Sy1   & 0.0290 & 73.01958  &  49.54583 &  112.5 & --\\
ESO 33-2        & Sy2   & 0.0181 & 73.99834  &--75.54056 & 1110.4 & --\\
4U 0517+17      & Sy1.5 & 0.0179 & 77.68958  &  16.49861 & 1145.4 &  19.8\\
Ark 120         & Sy1   & 0.0327 & 79.04784  & --0.15017 &  571.9 &
63.1 & $8.18 {+0.05 \atop -0.06}^g$ & R\\
IGR J05270-6631$^x$& QSO   & 0.978  & 81.56021  &--66.5125  &  631.7 &
323.4 & $8.4 \pm 0.5^a$ & LL/CL\\
PKS 0528+134$^x$& blazar& 2.060  & 82.73507  &  13.53199 &  917.6 & 157.3\\
NGC 2110        & Sy2   & 0.0078 & 88.04742  & --7.45622 &   20.5 & --
& $8.3 \pm 0.3^e$ & S\\
MCG+08-11-011   & Sy1.5 & 0.0205 & 88.72338  &  46.43934 &   50.8 & --
& $8.1 \pm 0.6^c$ & SO\\
IRAS 05589+2828 & Sy1   & 0.0330 & 90.54042  &  28.47139 & 1689.9 &  29.0 \\
SWIFT J0601.9-8636 & Sy2 &0.0064 & 91.41292  &--86.63111 &  177.8 &
31.9 & $7.9 \pm 0.5^n$ & KM\\
IGR J06117-6625 & Sy1.5 & 0.230  & 92.95208  &--66.40847 & 1122.2 &  97.2\\
Mrk 3           & Sy2   & 0.0135 & 93.90129  &  71.03748 &  814.9 &
55.4 & $8.7 \pm 0.3^e$ & S\\
IGR J06239-6052 & Sy2   & 0.0405 & 95.94004  &--60.97927 &  598.0 & --\\
IGR J06292+4858$^x$ & BLLac & 0.097  & 97.300    &  48.97389 &  --    & --\\
PKS 0637-752    & Sy1   & 0.651  & 98.94379  &--75.27133 & 1066.6 &
22.7 & $9.4 \pm 0.5^e$ & CL\\
Mrk 6           & Sy1.5 & 0.0188 & 103.0513  &  74.42689 &  819.4 &
109.7 & $8.2 \pm 0.5^n$ & KM\\
QSO B0716+714   & BLLac & 0.3    & 110.4727  &  71.34343 &  863.8 & 162.8\\
LEDA 96373      & Sy2   & 0.0294 & 111.6096  &--35.90583 &  180.9 &   5.6\\
IGR J07437-5137 & Sy2   & 0.025  & 115.9208  &--51.61694 &  947.3 &   --\\
IGR J07565-4139 & Sy2   & 0.021  & 119.0817  &--41.62836 & 1703.0 &  50.6\\
IGR J07597-3842 & Sy1   & 0.040  & 119.9242  &--38.73223 & 1288.1 &  31.8 
& $8.3 \pm 0.5^p$ & LL \\
ESO 209-12      & Sy1.5 & 0.0405 & 120.4900  &--49.77833 & 2288.6 & 117.7\\
PG 0804+761     & Sy1   & 0.10   & 122.7444  &  76.04514 &  694.9 &
44.9 & $8.84 {+0.05 \atop -0.06}^g$ & R\\
Fairall 1146    & Sy1.5 & 0.0316 & 129.6279  &--35.99306 & 1689.9 &  26.2\\
QSO B0836+710   & BLLac & 2.1720 & 130.3515  &  70.89506 &  754.4 &  29.7\\
IGR J09026-4812$^m$ & Sy1&0.039  & 135.6555  &--48.22608 & 3773.3 & 470.1\\
SWIFT J0917.2-6221& Sy1 & 0.0573 & 139.0392  &--62.32486 &  669.0 &  16.1\\
IGR J09253+6929 & Sy1   & 0.039  & 141.321   &  69.488   &  324.1 & --
& $7.6 \pm 0.5^a$ & LL/CL\\
Mrk 110         & NLS1  & 0.0353 & 141.3036  &  52.28625 &   54.2 & --
& $7.42 {+0.09 \atop -0.1}^g$ & R\\
IGR J09446-2636 & Sy1.5 & 0.1425 & 146.1500  &--26.60000 &  157.1 & --\\
NGC 2992       & Sy1   & 0.0077 & 146.4252  &--14.32639 &  383.3 &
66.7 & $7.7 \pm 0.3^e$ & S\\
MCG-05-23-016   & Sy2   & 0.0085 & 146.9173  &--30.94886 &  130.2 & --
& $6.3 \pm 0.5^c$ & SO\\
IGR J09523-6231 & Sy1.5 & 0.252  & 148.0854  &--62.54333 & 1059.1 & 111.1\\
NGC 3081        & Sy2   & 0.0080 & 149.8731  &--22.82628 &  255.3 & --
& $7.4 \pm 0.3^b$ & S\\
SWIFT J1009.3-4250& Sy2 & 0.033  & 152.4512  &--42.81222 &  437.4 & --\\
IGR J10147-6354 & Sy1.2 & 0.202  & 153.6750  &--63.89194 & 1182.9 &
104.6 & $8.6 \pm 0.5^a$ & LL/CL\\
NGC 3227        & Sy1.5 & 0.0039 & 155.8776  &  19.86492 &  142.0 & --
& $7.3 {+0.2 \atop -0.1}^h$ & K\\
NGC 3281        & Sy2   & 0.0107 & 157.9669  &--34.85369 &  211.3 &
41.9 & $8.0 \pm 0.5^n$ & KM\\
SWIFT J1038.8-4942&Sy1.5& 0.060  & 159.6875  &--49.78194 &  908.6 & 10.2\\
IGR J10404-4625 & Sy2   & 0.0237 & 160.0928  &--46.42353 &  673.6 & --\\
Mrk 421         & BLLac & 0.0300 & 166.1138  &  38.20883 &  826.1 &
482.4 & $8.3 \pm 0.3^e$ & S\\
IGR J11366-6002 & Sy2   & 0.014  & 174.1754  &--60.05217 & 2196.0 & 315.7 \\
NGC 3783        & Sy1   & 0.0097 & 174.7574  &--37.73853 &   22.9 & --
& $7.47 {+0.07 \atop -0.09}^g$ & R\\
IGR J12026-5349 & Sy2   & 0.028  & 180.6985  &--53.83547 & 1535.9 & 116.6\\
NGC 4051        & Sy1.5 & 0.0023 & 180.7901  &  44.53144 &  793.2 & --
& $6.3 \pm 0.2^g$ & R\\
NGC 4138        & Sy1.5 & 0.0030 & 182.3745  &  43.68500 &  798.4 &
31.4 & $6.8 \pm 0.5^n$ & KM\\
NGC 4151        & Sy1.5 & 0.0033 & 182.6364  &  39.40545 &  820.3 &
548.3 & $7.5 {+0.1 \atop -0.6}^h$ & K\\
NGC 4180        & AGN   & 0.0070 & 183.2627  &   7.03881 & 1016.8 &  72.1\\
Was 49          & Sy2   & 0.0610 & 183.5742  &  29.52872 &  931.4 & --\\
Mrk 766         & Sy1.5 & 0.0129 & 184.6110  &  29.81267 &  994.6 & --
& $6.5 \pm 0.3^i$ & S\\
NGC 4258        & Sy1.5 & 0.0015 & 184.7397  &  47.30397 &  816.2 &
4.6 & $7.59 \pm 0.01^i$ & M\\
4C 04.42        & BLLac & 0.9650 & 185.5940  &   4.22106 & 1277.5 & 232.6\\
Mrk 50          & Sy1   & 0.0234 & 185.8506  &   2.67911 & 1320.8 & 264.8\\
NGC 4388        & Sy2   & 0.0084 & 186.4455  &  12.66203 &  869.2 &
146.8 & $7.2 \pm 0.6^b$ & SO\\
NGC 4395        & Sy1.5 & 0.0011 & 186.4539  &  33.54661 & 1103.5 & --
& $4.7 {+0.3 \atop -0.7}^i$ & V\\
3C 273          & QSO   & 0.1583 & 187.2779  &   2.05239 & 2004.6 &
299.2 & $9.81 {+0.1 \atop -0.07}^k$ & R\\
NGC 4507        & Sy2   & 0.0118 & 188.9023  &--39.90925 &  389.3 & --
& $7.6 \pm 0.6^c$ & SO\\
SWIFT J1238.9-2720 & Sy2 &0.0250 & 189.7271  &--27.30778 &   71.3 & --
& $8.6 \pm 0.5^n$ & KM\\
IGR J12391-1612 & Sy2   & 0.0367 & 189.7762  &--16.17975 &  618.7 &
39.4 & $8.9 \pm 0.5^n$ & KM\\
NGC 4593        & Sy1   & 0.0090 & 189.9143  & --5.34425 & 1466.5 &
213.5 & $6.99 {+0.08 \atop -0.1}^h$ & R\\
IGR J12415-5750 & Sy1.5 & 0.0242 & 190.3575  &--57.83417 & 1653.1 &
101.6 & $8.0 \pm 0.5^a$ & LL/CL\\
PKS 1241-399    & QSO   & 0.1910 & 191.1223  &--40.21289 &  551.3 & --\\
ESO 323-32      & Sy1   & 0.0160 & 193.3348  &--41.63717 &  744.4 &  23.9\\
3C 279          & BLLac & 0.5362 & 194.0465  & --5.78931 & 1144.1 &
185.0 & $8.4 \pm 0.5^e$ & CL\\
IGR J13000+2529$^x$& AGN& --     & 195.0000  &  25.48333 &  937.1 & 241.0\\
Mrk 783         & Sy1.5 & 0.0672 & 195.7452  &  16.40763 &  728.2 &  11.2\\
IGR J13038+5348 & Sy1   & 0.0302 & 195.9975  &  53.79172 &  366.3 &
16.5 & $7.5 \pm 0.5^n$ & KM\\
NGC 4945        & Sy2   & 0.0019 & 196.3587  &--49.47083 & 1100.0 &
141.8 & $6.2 \pm 0.3^l$ & M\\
IGR J13057+2036 & AGN   & --     & 196.4273  &  20.58103 &  853.1 &  12.0\\
ESO 323-77      & Sy1   & 0.0150 & 196.6108  &--40.41389 &  920.1 &
65.3 & $7.4 \pm 0.6^c$ & LL\\
IGR J13091+1137 & Sy2   & 0.025  & 197.2733  &  11.63414 &  391.1 &
8.2 & $8.6 \pm 0.5^n$ & KM\\
IGR J13109-5552 & Sy1   & 0.104  & 197.6795  &--55.86991 & 1506.6 &  26.6\\
NGC 5033        & Sy1.5 & 0.0029 & 198.3650  &  36.59358 &  743.2 &   6.2\\
IGR J13149+4422 & Sy2   & 0.0366 & 198.8155  &  44.40750 &  421.4 &  56.8\\
Cen A           & Sy2   & 0.0018 & 201.3651  &--43.01911 & 1268.0 &
149.5 & $8.0 \pm 0.6^c$ & K\\
ESO 383-18      & Sy2   & 0.0124 & 203.3596  &--34.01631 &  756.8 & 191.6\\
MCG-06-30-015   & Sy1.2   & 0.0077 & 203.9741  &--34.29558 &  767.2 &
213.0 & $6.7 {+0.1 \atop -0.2}^i$ & S\\
NGC 5252        & Sy2   & 0.0230 & 204.5667  &   4.54236 &   58.9 & --
& $9.03 {+0.4 \atop -0.02}^l$ & K\\
Mrk 268         & Sy2   & 0.0399 & 205.2964  &  30.37811 &  728.7 & --\\
4U 1344-60      & Sy1.5 & 0.0129 & 206.8833  &--60.61000 & 1537.7 & 231.2\\
IC 4329A        & Sy1   & 0.0161 & 207.3304  &--30.30956 &  381.9 &
52.6 & $\sim 7^g$ & R\\
Circinus Galaxy & Sy2   & 0.0014 & 213.2871  &--65.34084 & 2272.1 &
201.3 & $6.04 {+0.07 \atop - 0.09}^l$ & M\\
NGC 5506        & Sy1.9 & 0.0062 & 213.3120  & --3.20750 &  102.9 &
88.5 & $6.7 \pm 0.7^b$ & SO\\
IGR J14175-4641 & Sy2   & 0.076  & 214.2664  &--46.69419 & 1268.8 &  21.4\\
NGC 5548        & Sy1.5 & 0.0172 & 214.4985  &  25.13706 &  200.9 &
42.4 & $7.82 \pm 0.02^h$ & R\\
RHS 39          & Sy1   & 0.0222 & 214.8425  &--26.64472 &  454.4 &
8.1 & $8.7 \pm 0.5^n$ & KM\\
H 1426+428      & BLLac & 0.1291 & 217.1358  &  42.67472 &  499.5 &
224.7 & $9.1 \pm 0.7^e$ & SB\\
IGR J14471-6414 & Sy1   & 0.053  & 221.6158  &--64.27319 & 1797.1 & 171.8\\
IGR J14471-6319 & Sy2   & 0.038  & 221.8120  &--63.28868 & 1842.9 & 230.3\\
IGR J14492-5535 & AGN   & --     & 222.3038  &--55.60578 & 1886.0 & 142.4\\
IGR J14515-5542 & Sy2   & 0.018  & 222.8880  &--55.67733 & 1869.2 & 144.0\\
IGR J14552-5133 & NLS1  & 0.016  & 223.8223  &--51.57102 & 2007.3 &   7.9
& $6.3 \pm 0.5^p$ & LL \\
IGR J14561-3738 & Sy2   & 0.024  & 224.0342  &--37.64803 & 2185.9 & 340.8\\
IGR J14579-4308 & Sy2   & 0.016  & 224.4296  &--43.13000 & 2012.9 & 333.4\\
Mrk 841         & Sy1.5 & 0.0364 & 226.0050  &  10.43782 &  131.3 &
6.7 & $8.5 \pm 0.7^c$ & R\\
ESO 328-36      & Sy1   & 0.0237 & 228.6958  &--40.35861 & 2200.9 & 384.8\\
IGR J15161-3827 & Sy2   & 0.0365 & 229.0375  &--38.44806 & 2181.2 & 372.1\\
NGC 5995        & Sy2   & 0.0252 & 237.1040  &--13.75778 & 1129.2 &  28.6\\
IGR J15539-6142 & Sy2   & 0.015  & 238.3967  &--61.68206 & 1754.0 & 101.0\\
IGR J16024-6107 & Sy2   & 0.0114 & 240.4517  &--61.14822 & 1792.4 &  84.3\\
IGR J16056-6110 & Sy1.5 & 0.052  & 241.4643  &--61.19525 & 1727.2 &  64.4\\
IGR J16119-6036 & Sy1   & 0.016  & 242.9642  &--60.63194 & 1789.6 &  60.9\\
IGR J16185-5928 & NLS1  & 0.035  & 244.6518  &--59.45482 & 2066.0 &  44.6
& $7.4 \pm 0.5^p$ & LL \\
IGR J16351-5806 & Sy2   & 0.009  & 248.8071  &--58.08047 & 2230.3 & 39.3\\
IGR J16385-2057 & NLS1  & 0.0269 & 249.6250  &--20.94389 & 1505.9 &  80.4\\
IGR J16426+6536 & NLS1  & 0.323  & 250.7670  &  65.54747 &   56.0 &
--  & $7.0 \pm 0.5^c$ & LL/CL\\
IGR J16482-3036 & Sy1   & 0.0313 & 252.0623  &--30.58502 & 2450.9 &  53.7\\
ESO 138-1       & Sy2   & 0.0091 & 252.8333  &--59.23389 & 1509.2 &  21.4\\
NGC 6221        & Sy2   & 0.0050 & 253.1942  &--59.21639 & 1481.0 &  21.4\\
NGC 6240        & Sy2   & 0.0245 & 253.2457  &   2.40047 &  300.5 &  67.7\\
Mrk 501         & BLLac & 0.0337 & 253.4676  &  39.76017 &  476.3 &
70.6 & $9.2 \pm 0.3^e$ & S\\
IGR J16558-5203 & Sy1   & 0.054  & 254.0234  &--52.06135 & 3065.7 & 249.6
& $7.9 \pm 0.5^p$ & LL \\
IGR J16562-3301 & BLLac & --     & 254.0701  &--33.03680 & 3669.2 & 111.4\\
NGC 6300        & Sy2   & 0.0037 & 259.2467  &--62.81972 &  237.9 &
8.9 & $5.5 \pm 0.4^c$ & X\\
IGR J17204-3554 & AGN   & --     & 260.1042  &--35.90000 & 6318.5 & 292.3\\
QSO B1730-130$^x$& QSO  & 0.9020 & 263.2613  &--13.08042 & 1080.1 &  33.1\\
GRS 1734-292    & Sy1   & 0.0214 & 264.3681  &--29.13403 & 9122.1 & 665.7
& $8.9 \pm 0.7^q$ & SO \\
IGR J17418-1212 & Sy1   & 0.0372 & 265.4625  &--12.19611 & 1300.1 &  45.5\\
IGR J17488-3253 & Sy1   & 0.020  & 267.2297  &--32.91449 & 6159.4 & 410.4\\
IGR J17513-2011 & Sy2   & 0.047  & 267.8068  &--20.20405 & 6220.1 & 268.7
& $6.0 \pm 0.5^p$ & LL \\
IGR J18027-1455 & Sy1   & 0.0034 & 270.6974  &--14.91522 & 2463.8 & 173.0 \\
IGR J18244-5622 & Sy2   & 0.0169 & 276.0812 & --56.36909 &  103.3 &   4.4\\
IGR J18249-3243 & Sy1   & 0.355  & 276.2361 & --32.71661 & 5126.8 & 208.7\\
IGR J18259-0706 & Sy1?  & --     & 276.48958&  --7.17264 & 2009.5 & 226.1\\
PKS 1830-211    & BLLac & 2.5070 & 278.4162 & --21.06106 & 2775.4 & 111.2\\
3C 382          & Sy1   & 0.0579 & 278.7641 &   32.69635 &   15.2 &
0.7 & $9.2 \pm 0.5^n$ & KM\\
ESO 103-35      & Sy2   & 0.0133 & 279.5846 & --65.42805 &   28.5 & --
& $7.1 \pm 0.6^c$ & X\\
3C 390.3        & Sy1   & 0.0561 & 280.5374 &   79.77142 &  448.0 &
33.2 & $8.46 {+0.09 \atop -0.1}^g$ & R\\
ESO 140-43$^*$  & Sy1   & 0.0141 & 281.2917 & --62.35583 &  378.0 & 27.7\\
IGR J18559+1535 & Sy1   & 0.084  & 284.0000 &   15.63694 & 2594.9 &  24.7\\
ESO 141-55$^*$  & Sy1   & 0.0366 & 290.3092 & --58.67083 &   378.0
& 79.9 & $7.1 \pm 0.6^c$ & SO\\
1RXS J192450.8-29143& BLLac & 0.3520 & 291.2127 &--29.24170&955.7 &  28.6\\
1H 1934-063     & Sy1   & 0.0106 & 294.3879 &  --6.21806 &  654.2 &
2.1 & $7.9 \pm 0.6^c$ & SO\\
IGR J19405-3016 & Sy1   & 0.052  & 295.0631 & --30.26347 &  930.3 &  22.3\\
NGC 6814        & Sy1.5 & 0.0052 & 295.6683 & --10.32333 &  438.5 &
7.4 & $7.1 \pm 0.2^h$ & CL\\
IGR J19473+4452 & Sy2   & 0.0539 & 296.8307 &   44.82845 & 1194.3 &   1.0\\

3C 403          & Sy2   & 0.0590 & 298.0617 &    2.50778 &  490.2 & --\\
QSO B1957+405   & Sy2   & 0.0561 & 299.8682 &   40.73386 & 2258.1 &
69.5 & $9.4 \pm 0.1^l$ & K\\
1ES 1959+650$^x$    & BLLac & 0.048  & 299.9994 &   65.14851 &   11.2 & --
& $8.1 \pm 0.3^e$ & S\\
ESO 399-20      & NLS1  & 0.0250 & 301.7383 & --34.54833 &  857.2 &  24.7\\
IGR J20187+4041 & Sy2   & 0.0144$^r$& 304.6606 &   40.68344 & 2851.0 & 442.7\\
IGR J20286+2544 & Sy2   & 0.013  & 307.1462 &   25.73361 &  840.2 &   4.1\\
4C 74.26$^x$    & QSO   & 0.1040 & 310.6549 &   75.13403 &   77.1 &
-- & $9.6 \pm 0.5^e$ & CL\\
Mrk 509         & Sy1.2 & 0.0344 & 311.0406 & --10.72348 &   71.7 &
57.3 & $8.16 {+0.03 \atop -0.04}^g$ & R\\
S5 2116+81      & Sy1   & 0.086  & 318.5021 &   82.07975 &  176.5 &
-- & $8.8 \pm 0.5^n$ & KM\\
IGR J21178+5139 & AGN   & --     & 319.4468 &   51.64823 & 1261.9 & 171.3\\
IGR J21247+5058 & Sy1   & 0.020  & 321.1640 &   50.97329 & 1392.7 & 174.7\\
IGR J21272+4241$^x$ & Sy1.5 & 0.316  & 321.7917 &   42.69194 &  836.3 & 189.8\\
IGR J21277+5656 & Sy1   & 0.0144 & 321.9373 &   56.94436 &  984.5 & 166.6\\
RX J2135.9+4728 & Sy1   & 0.0252 & 323.9766 &   47.47453 & 1213.4 & 136.3\\
PKS 2149-306    & FSRQ  &  2.345 & 327.9813 & --30.46492 &  258.6 &  47.1\\ 
NGC 7172        & Sy2   & 0.0086 & 330.5071 & --31.87167 &  346.5 &
68.8 & $7.7 \pm 0.6^c$ & SO\\
BL Lac          & BLLac & 0.0686 & 330.6804 &   42.27778 &  738.1 &
-- & $8.2 \pm 0.7^e$ & SB\\
IGR J22292+6647 & Sy1$^o$   & 0.113$^o$ & 337.3062 &   66.78106 & 1678.9 &   9.3\\
NGC 7314        & Sy1   & 0.0048 & 338.9419 & --26.05047 &  546.8 &
1.7 & $6.0 \pm 0.5^c$ & S\\
Mrk 915         & Sy1   & 0.0241 & 339.1938 & --12.54517 &  609.0 &  12.8\\
IGR J22517+2217 & BLLac & 3.668  & 342.9280 &   22.29900 &  251.3 &   5.8\\
3C 454.3        & BLLac & 0.8590 & 343.4906 &   16.14822 &  207.1 &
36.3 & $9.2 \pm 0.7^e$ & CL\\
1H 2251-179     & Sy1   & 0.0640 & 343.5245 & --17.58203 &  579.2 &
314.6 & $<6.9^c$ & W\\
NGC 7469        & Sy1   & 0.0163 & 345.8156 &    8.87386 &  139.2 &
-- & $7.09 \pm 0.05^g$ & R\\
MCG-02-58-022   & Sy1.5 & 0.0469 & 346.1812 &  --8.68572 &  624.5 &
28.2 & $7.1 \pm 0.6^c$ & SO\\
NGC 7603$^x$    & Sy1.5 & 0.0295 & 349.7359 &    0.24347 &  104.5 &
-- & $8.1 \pm 0.3^e$ & S\\
IGR J23206+6431 & Sy1   & 0.0732 & 350.15   &   64.52    & 3888.2 & 349.8\\
IGR J23308+7120 & Sy2   & 0.037  & 352.6552 &   71.37911 & 1965.1 &  --\\
IGR J23524+5842 & Sy2   & 0.164  & 358.0917 &   58.75908 & 4076.7 & 361.1\\


\end{longtable}
$^{*}$ Beckmann, Petry, \& Weidenspointner 2007b, $^a$ Masetti et
al. 2009, $^b$ Bian \& Gu 2007, $^c$ Middleton, Done \& Schurch 2008,
$^d$ Kaspi et al. 2000, $^e$ Woo \& Urry 2002, $^f$ Merloni, Heinz \&
Di Matteo 2003, $^g$ Peterson et al. 2004, $^h$ Hicks \& Malkan 2008,
$^i$ Uttley \& McHardy 2005, $^k$ Paltani \& T\"urler 2005, $^l$
Graham 2008, $^m$ Zurita-Heras et al. 2009, $^n$ Winter et al. 2009,
$^o$ Butler et al. 2009, $^p$ Masetti et al. 2006, $^q$ see Appendix
\ref{section:singlesource}, $^r$ Goncalves et al. 2009, $^x$ not detected in the data set
presented here
\begin{longtable}{lrccccc}
\caption{Spectral fit results for IBIS/ISGRI
 data\label{fitresults}.}\\
\hline\hline
Name &  ISGRI & $N_{\rm H}$ & $f_{20-40 \rm \, keV}$ & $f_{40-100 \rm \,
keV}$ & $\Gamma_{\rm ISGRI}^+$ & $\log L_{20 - 100 \rm \, keV}$\\
 &  18--60 keV $[\sigma]$ & $[10^{22} \rm \, cm^{-2}]$ &  $[10^{-11} \rm \, erg \, cm^{-2}
\, s^{-1}]$ &  $[10^{-11}
\rm \, erg \, cm^{-2} \, s^{-1}]$  & &  $[\rm \, erg \, s^{-1}]$ \\ 
\hline
\endfirsthead
\caption{continued.}\\
\hline\hline
Name &  ISGRI &  $N_{\rm H}$ & $f_{20-40 \rm \, keV}$ & $f_{40-100 \rm \, keV}$ & $\Gamma_{\rm ISGRI}^+$ & $\log L_{20 - 100 \rm \, keV}$\\
 &  18--60 keV $[\sigma]$ & $[10^{22} \rm \, cm^{-2}]$ &  $[10^{-11}
\rm \, erg \, cm^{-2} \, s^{-1}]$ &  $[10^{-11}
\rm \, erg \, cm^{-2} \, s^{-1}]$ & &  $[\rm \, erg \, s^{-1}]$\\ 
\hline
\endhead
\hline
\endfoot
IGR J00254+6822 & 7.5  & $40^c$   & 0.6 & 0.8 & $2.1 \pm 0.5$&
42.67  \\
IGR J00335+6126 & 5.9 & $0.5^*$ & 0.3 & 0.9 & $1.1 \pm 0.2$ & 44.50 \\
1ES 0033+59.5   & 12.6 & $0.36^a$  & 0.8 & 0.3 & $3.6 {+0.4 \atop
 -0.3}$ &44.36 \\  
Mrk 348         & 14.6 & $30^b$   & 4.2 & 6.6 & C   &43.74 \\ 
NGC 526A        & 4.3 & $1.6^b$ & 2.2 & 2.9 & 2 & 43.62 \\
ESO 297-18      & 5.2 & $42^i$  & 3.1 & 6.8 & $1.5 \pm 0.3$ &44.10 \\
IGR J01528-0326 & 8.5 & $14^c$  & 0.9 & 2.3 & $1.2 {+0.2 \atop -0.5}$ &
43.30\\
NGC 788         & 23.4 & $<0.02^a$ & 2.5 & 3.9 & $1.8 \pm 0.1$         & 43.42 \\
Mrk 590         & 4.5 & 0.03 & 0.5 & 0.6 & 2 & 43.23\\
IGR J02097+5222 & 7.7 & $0.03^i$ & 1.2 & 2.0 & $1.7 {+0.3 \atop -0.2}$ &44.25 \\
SWIFT J0216.3+5128 & 5.1 & $1.27^f$ & 0.8 & 0.8 & $2.3 {+1.2 \atop -1.0}$ &43.48 \\ 
Mrk 1040       & 3.4 & $0.067^b$ & 2.3 & 3.0 & 2 & 43.52 \\
IGR J02343+3229 & 3.5 & $2.2^d$ & 1.4 & 1.8 & 2 &                       43.27\\
NGC 985         & 5.8 & $0.6^b$ & 0.9 & 1.3 & $2.0 {+0.4 \atop -0.3}$ & 43.99 \\
NGC 1052        & 5.6 & $0.041^b$ & 0.8 & 1.5 & $1.6 {+0.8 \atop -0.4}$                    & 42.11 \\
RBS 345    &  5.0 & -- & 0.9 & 1.9 & $1.3 {+0.6 \atop -0.4}$ & 44.49 \\ 
NGC 1068        &  7.3 & $>150^a$& 1.2 & 1.3 & $2.3 {+0.4 \atop -0.3}$ & 43.68\\
QSO B0241+62    & 16.3 & $1.5^a$ & 2.0 & 3.0 & $1.8 \pm 0.1$ & 44.37 \\
IGR J02466--4222 &  3.3 & $1$   & 1.0 & 1.4 & 2 & 44.45 \\
IGR J02501+5440 &  5.1 & --     & 0.6 & 1.5 & $1.2 \pm 0.3$ & 43.03 \\ 
MCG-02-08-014   &  6.3 & --      & 1.3 & 0.9 & $2.7 {+0.6 \atop -0.5}$  & 43.15 \\
NGC 1142        & 16.2 & $45^b$  & 2.7 & 4.2 & $1.8  \pm 0.1$ & 44.11 \\
QSO B0309+411   &  4.0 & $<0.1^*$& 3.0 & 3.9 & 2 & 45.53 \\
NGC 1275        & 15.7 & $3.75^a$& 2.2 & 0.4 & $3.7 {+0.3 \atop -0.2}$ & 43.18\\
1H 0323+342     &  5.1 & $0.1^b$ & 0.7 & 2.4 & $1.0 {+0.4 \atop -0.7}$ &
44.43 \\
IGR J03334+3718 &  6.9 & --     & 1.0 & 1.5 & $1.9 \pm 0.3$ & 44.25\\
NGC 1365        &  5.1 & $44^b$  & 1.5 & 2.2 & $1.8 \pm 0.6$ & 42.38 \\
IGR J03532-6829 &  7.3 & $0.05^*$ & 1.4 & 0.6 & $3.5 {+0.7 \atop -0.6}$ & 44.63 \\
3C 111          & 13.7 & $0.63^a$& 5.3 & 6.9 & $2.0 \pm 0.2$ & 44.83 \\
3C 120          & 22.0 & $0.2^i$ & 3.0 & 4.4 & $1.8 \pm 0.1$ & 44.27 \\
UGC 3142        & 17.3 & $1.4^x$ & 2.8 & 3.7 & $2.0 \pm 0.1$ & 44.84 \\
LEDA 168563     &  5.4 & $<0.22^i$& 2.2 & 3.1 & $1.9 {+0.5 \atop -0.4}$&44.00 \\ 
ESO 33-2        & 15.5 & $0.1^b$ & 1.3 & 2.0 & $1.8 \pm 0.2$ & 43.38 \\
4U 0517+17      & 24.5 & $0.1^b$ & 3.0 & 3.6 & $2.13 {+0.09 \atop -0.08}$ &
43.68 \\
Ark 120         & 15.6 & $<0.1^*$ & 2.7 & 3.4 & $2.1 \pm 0.1$ & 44.18 \\
NGC 2110        &  4.9 & $4.3^*$& 7.0 & 9.3 & 2 & 43.34 \\
MCG+08-11-011   &  7.8 & $0.183^b$& 5.2 & 6.7 & $2.0 \pm 0.4$& 44.05 \\
IRAS 05589+2828 & 15.4 & $<0.04^i$& 1.4 & 2.8 & $1.5 {+0.2 \atop -0.1}$ &
44.02 \\
SWIFT J0601.9-8636 & 6.5 & $5.6^i$&1.3 & 1.0 & $2.6 {+0.7 \atop -0.6}$ &
42.32 \\
IGR J06117-6625 & 17.8 & $0.048^b$& 1.9 & 2.6 & $1.96 \pm 0.16$ & 45.85 \\
Mrk 3           & 26.7 & $110^a$  & 3.7 & 6.4 & $1.73 \pm 0.07$ & 43.61  \\
IGR J06239-6052 &  3.6 & $20^h$& 0.5 & 0.7 & 2 & 43.65 \\
PKS 0637-752    &  5.4 & $0.035^b$& 0.7 & 1.1 &  $1.7 \pm 0.6$ & 46.45\\
Mrk 6           & 15.1 & $10^a$& 1.9 & 2.3 & $2.1 \pm 0.2$ & 43.53 \\
QSO B0716+714   &  4.6 & $<0.01^a$& 0.3 & 0.3 & 2 & 45.24 \\
LEDA 96373      &  5.3 & -- & 1.4 & 2.6 & $1.5 {+0.7 \atop -0.5}$ & 43.89 \\
IGR J07437-5137&  3.5 & -- & 0.4 & 0.5 & 2 & 43.13 \\
IGR J07565-4139 &  6.2 & $1.1^b$& 0.7 & 0.6 & $2.7 {+0.6 \atop -0.4}$ & 43.12 \\
IGR J07597-3842 & 14.2 & $0.05^b$& 1.8 & 1.9 & $2.3 \pm 0.2$ & 44.14 \\
ESO 209-12      & 12.2 & $0.1^b$& 0.9 & 1.4 & $1.8 {+0.1 \atop -0.2}$ & 43.93 \\
PG 0804+761     &  6.3 & $0.023^b$ & 1.1 & 1.5 & $2.0 \pm 0.5$ & 44.81\\
Fairall 1146    &  9.9 & $0.1^b$& 0.9 & 1.4 & $1.8 {+0.3 \atop -0.2}$&43.73\\
QSO B0836+710   & 15.4 & $0.11^a$& 2.1 & 4.1 & $1.5 {+0.2 \atop -0.1}$ & 47.93\\
IGR J09026-4812  & 17.8 & $0.9^w$ & 0.9 & 1.4 & $1.9 \pm 0.1$ & 43.91\\
SWIFT J0917.2-6221&5.8 & $0.5^f$& 0.8 & 1.1 & $1.9 {+0.5 \atop -0.6}$   & 44.18 \\
IGR J09253+6929 &  7.0 & $8^d$& 1.8 & 2.6 & $1.9 \pm 0.4$ & 44.19\\
Mrk 110        &  3.6 & $0.019^b$& 2.2 & 3.0 & 2 & 44.18 \\
IGR J09446-2636&  3.2 & $0.1^b$& 1.3 & 1.7 & 2 & 45.60 \\
NGC 2992        & 15.8 & $0.1^l$& 2.8 & 4.1 & $1.9 \pm 0.1$ & 42.95\\
MCG-05-23-016   & 14.6 & $1.6^a$ & 5.6 & 8.1 & $1.9 \pm 0.1$ & 43.34\\
IGR J09523-6231 &  5.3 & $8^d$ & 0.6 & 0.7 & $2.2 {+0.6 \atop -0.5}$ & 45.41 \\
NGC 3081        &  9.7 & $66^m$ & 2.6 & 4.1 & $1.8 \pm 0.2$ & 42.97 \\
SWIFT J1009.3-4250&6.9 &  $30^f$ & 1.5 & 2.0 & $2.0 \pm 0.4$ & 43.95\\
IGR J10147-6354&  3.0 & $2^d$  & 0.5 & 0.7 & 2 & 45.18\\ 
NGC 3227        & 16.6 & $6.8^b$ & 5.6 & 7.1 & $2.0 \pm 0.1$ & 42.63\\
NGC 3281        &  7.5 & $151^b$ & 1.8 & 2.7 & $1.9 \pm 0.3$ & 43.06\\
SWIFT J1038.8-4942&5.3 & $0.6^f$ & 0.6 & 1.4 & $1.3 \pm 0.4$ & 44.21 \\
IGR J10404-4625 &  6.9 & $3^f$ & 1.2 & 1.4 & $2.2 {+0.4 \atop -0.3}$ & 43.54\\
Mrk 421         &173.8 & $0.08^i$ & 22.4 & 20.6 & $2.45 {0.03 \atop -0.02}$ & 44.92\\
IGR J11366-6002 &  7.6 & $0.35^o$ & 0.6 & 0.7 & $2.2 {+0.5 \atop -0.4}$ & 42.73 \\
NGC 3783       &  6.1 & $0.08^*$ & 6.0 & 8.4 & $1.9 \pm 0.4$ & 43.48\\
IGR J12026-5349 & 14.8 & $2.2^a$ & 1.6 & 2.2 & $2.0 \pm 0.1$ & 43.84\\
NGC 4051        & 11.7 &  $<0.02^*$ & 1.8 & 1.5 & $2.1 \pm 0.2$ & 41.58 \\
NGC 4138        &  9.0 & $8^b$ & 1.2 & 1.9 & $1.8 \pm 0.3$ & 41.79 \\
NGC 4151       & 205.7 & $6.9^a$ & 24.0 & 32.3 & C & 43.13\\
NGC 4180        &  9.4 & -- & 0.9 & 1.8 & $1.6 {+0.3 \atop -0.2}$ & 42.46 \\
Was 49         &  3.8 & $10^b$ & 0.6 & 0.8 & 2 & 44.12\\
Mrk 766         &  6.9 & $0.8^a$ & 1.0 & 1.5 & $1.8 \pm 0.3$ & 42.95\\
NGC 4258        &  6.6 & $8.7^b$ & 1.1 & 1.1 & $2.3 \pm 0.4$ & 41.04\\
4C 04.42        &  7.7 & $0.1^b$  & 0.7 & 1.8 & $1.2 \pm 0.2$ & 46.83\\
Mrk 50          &  6.2 & $0.018^b$ & 0.5& 0.4 & $2.6 {+0.7 \atop -0.6}$ & 43.09\\
NGC 4388        & 78.1 & $27^a$ & 9.8 & 15.2 & C & 43.59\\
NGC 4395        &  8.8 & $0.15^a$ & 1.5 & 1.6 & $2.3 {+0.4 \atop -0.3}$ & 40.85\\
3C 273          & 78.0 & $0.5^a$ & 7.5 & 10.6 & $1.92 \pm 0.03$ & 46.09\\
NGC 4507        & 30.9 & $29^a$ & 6.3 &10.0 & C & 43.70\\
SWIFT J1238.9-2720&6.3 & $60^f$ & 2.7 & 5.3 & $1.6 {+0.4 \atop -0.3}$ & 44.06\\
IGR J12391-1612 &  9.9 & $3^f$ & 1.7 & 2.2 & $2.0 \pm 0.2$ & 44.08\\
NGC 4593        & 33.0 & $0.02^a$ & 3.2 & 4.1 & C & 43.08\\
IGR J12415-5750 &  6.7 & $<0.11^b$ & 0.8 & 1.4 & $1.7 {+0.2 \atop -0.3}$ & 43.45\\
PKS 1241-399    &  3.5 & -- & 0.7 & 0.9 & 2 & 45.23\\
ESO 323-32      &  5.6 & $7^f$ & 0.8 & 1.5 & $1.6 {+0.4 \atop -0.3}$ & 43.13\\
3C 279          &  8.6 & $\lae 0.13^a$ & 0.9 & 1.7 & $1.6 \pm 0.2$ & 46.40\\
Mrk 783        &  4.7 & $0.046^b$ & 0.9 & 1.2 & 2  & 44.36\\
IGR J13038+5348 &  4.5 & $<0.03^*$ & 1.0 & 1.3 & 2 & 43.67\\
NGC 4945        & 78.7 & $400^a$ & 9.9 & 18.4 & C & 42.35\\
IGR J13057+2036 &  3.1 & --  & 0.4 & 0.6 &  2   & \\
ESO 323-77      &  9.0 & $55^a$ & 1.1 & 2.1 & $1.5 {+0.3 \atop -0.2}$ & 43.21\\
IGR J13091+1137 &  7.5 & $90^a$ & 1.6 & 2.5 & $2.1 \pm 0.3$           & 43.76\\
IGR J13109-5552 &  8.7 & $<0.1^p$  & 0.9 & 1.2 & $2.0 {+0.3 \atop -0.2}$ & 44.78\\
NGC 5033        &  4.2 & $0.03^b$ & 0.6 & 0.8 & 2  & 41.43\\
IGR J13149+4422 & 4.7 & $5^d$ & 0.8 & 1.1 & 2 & 43.77\\
Cen A           &247.2 & $12.5^a$ & 28.1 & 43.4 & $1.82 \pm 0.01$ & 42.71\\
ESO 383-18      &  6.5 & $17^x$ & 0.9 & 1.2 & $1.9 {+0.4 \atop -0.3}$ & 42.86\\
MCG-06-30-015   & 21.1 & $0.03^b$ & 2.3 & 1.7 & $2.4 \pm 0.1$ & 42.74\\
NGC 5252        &  3.0 & $0.68^b$  & 2.0 & 2.7 & 2 & 43.76\\
Mrk 268         &  5.7 & --  & 0.9 & 0.7 & $2.5 \pm 0.6$ & 43.78\\
4U 1344-60      & 35.2 & $5^a$ & 3.5 & 4.6 & C & 43.48\\
IC 4329A        & 58.5 & $0.42^a$ & 8.8 & 12.3 & C & 44.09\\
Circinus Galaxy &120.0 & $360^a$  & 10.9 & 10.1 & C & 41.96\\
NGC 5506        & 27.8 & $3.4^a$  & 7.0 & 6.2 & $2.1 \pm 0.1$ & 43.16\\
IGR J14175-4641 &  8.0 & --       & 0.8 & 1.3 & $1.8 \pm 0.3$& 44.48\\
NGC 5548        &  3.1 & $0.51^a$  & 1.0 & 1.3 & 2 & 43.19\\
RHS 39          &  7.7 & $<0.05^a$ & 1.6 & 2.2 & $2.0 \pm 0.3$& 43.63\\
H 1426+428      &  4.6 & $<0.02^*$ & 0.6 & 0.8 & 2 & 44.80\\
IGR J14471-6414 &  7.2 & $<0.1^*$ & 0.5 & 0.8 & $1.8 \pm 0.3$    & 43.95\\
IGR J14471-6319 &  6.4 & $2^f$ & 0.6 & 0.8 & $2.0 {+0.5 \atop -0.6}$    & 43.67\\
IGR J14492-5535 &  8.6 & $12^q$ & 0.9 & 1.4 & $1.7 \pm 0.3$ & \\
IGR J14515-5542 &  9.4 & $0.4^f$ & 0.8 & 1.4 & $1.6 \pm 0.2$  & 43.19\\
IGR J14552-5133 &  8.5 & $0.1^b$ & 0.6 & 1.1 & $1.6 {+0.4 \atop -0.2}$  & 42.98\\
IGR J14561-3738 &  9.4 & $>100^q$ & 0.6 & 1.2 & $1.7 {+0.2 \atop -0.4}$ & 43.37\\
IGR J14579-4308 & 15.4 & $20^d$  & 1.1 & 1.4 & $2.0 \pm 0.2$  & 43.16\\
Mrk 841         &  3.2 & $0.21$ & 1.0 & 1.4 & 2 & 43.87\\
ESO 328-36      &  5.7 & -- & 0.4 & 0.4 & $2.5 \pm 0.6$            & 43.00\\
IGR J15161-3827 & 12.3 & -- & 1.0 & 1.7 & $1.7 \pm 0.2$            & 43.90\\
NGC 5995        & 16.4 & $0.7^*$ & 2.2 & 1.5 & $2.8 \pm 0.2$    & 43.74\\
IGR J15539-6142 & 4.6 & $18^f$ & 0.5 & 0.7 & 2 & 42.78\\
IGR J16024-6107 &  3.7 & $<0.1^o$ & 0.3 & 0.4 & 2             & 42.46\\
IGR J16056-6110 &  3.8 & $<1^c$ & 0.3 & 0.5 & 2 & 43.72\\
IGR J16119-6036 &  9.3 & $0.1^b$ & 0.9 & 1.4 & $1.8 {+0.3 \atop -0.2}$ & 43.11\\
IGR J16185-5928 &  6.1 & $<0.1^e$ & 0.5  & 0.9 & $1.6 \pm 0.3$  & 43.57 \\
IGR J16351-5806 & 10.2 & $<0.1^s$ & 0.8 & 1.6 & $1.5 \pm 0.2$ & 42.64\\
IGR J16385-2057 &  7.2 & $0.21^d$ & 0.8 & 0.5 & $3.1 \pm 0.4$           & 43.35\\
IGR J16426+6536 &  4.4 &    --    & 2.6 & 3.4 & 2             & 46.32\\
IGR J16482-3036 & 27.5 & $0.13^f$ & 2.5 & 3.1 & $2.1 \pm 0.1$  & 44.10\\
ESO 138-1       &  3.7 & $150^b$ & 0.4 & 0.6 & 2 & 42.26\\
NGC 6221       &  4.7 & $1^a$ & 0.5 & 1.0 & 2 & 41.93\\
NGC 6240        & 12.1 & $2^b$ & 2.1 & 3.6 & $1.7 \pm 0.2$          & 43.89\\
Mrk 501         & 12.4 & $0.013^b$ & 2.5 & 1.7 & $2.8 \pm 0.3$& 44.05\\
IGR J16558-5203 & 16.5 & $0.011^b$ & 1.1 & 1.5 & $2.0 \pm 0.1$     & 44.25\\
IGR J16562-3301 &  7.0 & $0.2^f$ & 1.0 & 2.3 & $1.32 {+0.16 \atop -0.07}$ & \\
NGC 6300        & 10.1 & $22^a$ & 3.1 & 3.6 & $2.2 \pm 0.2$ & 42.30\\
IGR J17204-3554 &  6.4 & $12^b$ & 0.3 & 0.4 & $2.2 \pm 0.5$ & \\
GRS 1734-292    &106.4 & $3.7^a$ & 4.3 & 4.4 & C         & 43.96\\
IGR J17418-1212 &  6.6 & $0.1^b$  & 0.8 & 1.5 & $1.5 \pm 0.2$ & 43.85\\
IGR J17488-3253 & 27.5 & $0.7^*$ & 1.1 & 1.6 & C       & 43.39\\
IGR J17513-2011 & 16.0 & $0.6^*$ & 0.8 & 1.1 & $1.95 \pm 0.12$ & 44.00  \\ 
IGR J18027-1455 & 15.2 & $19.0^a$ & 1.2 & 1.7 & $1.9 \pm 0.1$ & 41.86\\
IGR J18244-5622 &  3.0 & $14^f$ & 1.3 & 1.7 & 2                    & 43.28\\
IGR J18249-3243 &  7.5 & $<0.1^n$ & 0.5 & 0.7 & $1.9 \pm 0.4$ & 45.72\\
IGR J18259-0706 &  7.8 & $0.6^f$ & 0.6 & 0.8 & $2.1 \pm 0.3$ & \\
PKS 1830-211    & 25.2 & $\lae 0.7^a$ & 1.9 & 3.7 & $1.49 {+0.05 \atop
 -0.07}$ & 48.19\\ 
3C 382          &  3.0 & $0.88^b$ & 3.8 & 5.0 & 2 & 44.85\\
ESO 103-35     &  4.9 & $19^a$ & 3.0 & 4.0 & 2  & 43.44\\
3C 390.3        & 14.3 & $<0.1^a$ & 2.3 & 3.9& $1.7 \pm 0.1$  & 44.67\\
ESO 140-43      & 14.0 & $1.8^x$ & 1.7 & 2.6 & $1.7 \pm 0.2$ & 43.40\\
IGR J18559+1535 & 12.0 & $0.7^d$ & 0.9 & 1.3 & $2.0 \pm 0.2$  & 44.59\\
ESO 141-55      & 19.0 & $0.004^t$ & 1.9 & 2.8 & $1.9 \pm 0.4$ & 44.30\\
1RXS J192450.8-29143& 4.6 & $0.088^b$ & 0.5 & 0.7 & 2 & 45.70\\
1H 1934-063     &  9.2 & $0.1^b$ & 1.3 & 0.7 & $3.1 {+0.4 \atop -0.3}$  & 42.71\\
IGR J19405-3016 & 11.8 & $<0.1^o$ & 1.3 & 1.9 & $1.8 \pm 0.2$ & 44.30\\
NGC 6814        & 16.2 & $<0.05^a$ & 3.0 & 4.6 & $1.8 \pm 0.1 $  & 42.65\\
IGR J19473+4452 &  7.1 & $11^a$ & 1.1 & 2.0 & $1.3 \pm 0.3$   & 44.33\\
3C 403          &  4.5 & $45^b$ & 0.9 & 1.2 & 2   & 44.23\\
QSO B1957+405   & 42.6 & $20^a$ & 3.7 & 5.0 & $1.97 \pm 0.05$   & 44.81\\
ESO 399-20     &  4.3 & $0.048^b$ & 0.6 & 0.9 & 2 & 43.33\\
IGR J20187+4041 & 12.2 & $6.1^b$ & 0.9 & 1.3 & $1.9 {+0.1 \atop -0.2}$ & 42.99\\
IGR J20286+2544 & 10.4 & $42^f$ & 1.8 & 3.3 & $1.6 \pm 0.2$  & 43.28\\
Mrk 509         & 11.2 & $<0.01^a$ & 3.8 & 4.1 & $2.3 \pm 0.3$& 44.34\\
S5 2116+81     &  4.4 & $<0.1^a$ & 1.3 & 1.8 & 2 & 44.76\\
IGR J21178+5139 &  7.7 & $2^f$ & 0.8 & 1.5 & $1.5 \pm 0.2$    & \\
IGR J21247+5058 & 47.3 & $0.6^k$ & 4.7 & 7.3 & C & 44.04\\
IGR J21277+5656 & 13.6 & $0.1^b$ & 1.6 & 1.8 & $2.2 \pm 0.2$            & 43.20\\
RX J2135.9+4728 &  9.2 & $0.4^i$ & 0.9 & 1.6 & $1.7 \pm 0.2$            & 43.55\\
PKS 2149-306    &  4.6 & 0.03 & 0.9 & 1.2 & 2 & 47.98\\
NGC 7172        & 19.4 & $9.0^a$ & 3.6 & 4.7 & $2.0 \pm 0.1$   & 43.13\\
BL Lac          &  5.5 & $0.3^b$ & 0.7 & 1.2 & $1.8 {+0.4 \atop -0.3}$ & 44.34\\
IGR J22292+6647 &  5.7 & $0.2^*$ & 0.5 & 0.6 & $2.3 \pm 0.5$ & 44.57\\
NGC 7314        & 12.9 & $0.122^b$ & 2.1 & 2.5 & $2.2 \pm 0.3$ & 42.37 \\
Mrk 915         &  4.5 & $<0.1^*$ & 0.6 & 0.8 & 2    & 43.28\\
IGR J22517+2217 &  6.6 & $3^v$ & 1.6 & 3.4 & $1.4 \pm 0.4$ & 48.43\\
3C 454.3        & 30.8 & $0.5^b$ & 7.1 & 13.2 & $1.58 \pm 0.06$ &  47.76\\
1H 2251-179     & 22.9 & $<0.19^a$ & 3.3 & 3.8 & $2.2 \pm 0.1$  & 44.85\\
NGC 7469        &  6.1 & $0.061^b$ & 1.8 & 2.3 & $2.1 \pm 0.4$  & 44.39\\
MCG-02-58-022   & 15.0 & $<0.08^a$ & 2.0 & 1.4 & $2.8 \pm 0.2$   & 44.26\\
IGR J23206+6431 &  7.7 & $0.6^*$ & 0.4 & 0.7 & $1.5 {+0.5 \atop -0.3}$  & 44.15\\
IGR J23308+7120 &  3.7 & $6^o$ & 0.3 & 0.4 & 2 & 43.39\\
IGR J23524+5842 &  6.7 & $6^d$ & 0.4 & 0.5 & $2.0 \pm 0.5$ & 44.79\\
\end{longtable}

$^+$ `C' indicates that a more complex model is
 required to fit the data (see Table\ref{complexISGRIspec}),
$^*$ This work, $^a$ Beckmann et al. 2006 and references therein, $^b$
 Bodaghee et al. 2007 and ref. therein, $^c$ Landi et al. 2007b, $^d$ Rodriguez et al. 2008, $^e$ Malizia et al. 2008, $^f$
 Malizia et al. 2007, $^g$ Beckmann et al. 2007b, $^h$ Revnivtsev et al. 2007, $^i$ Winter et
 al. 2008, $^k$ Ricci et al. 2009a, $^l$ Beckmann et al. 2007a, $^m$
 Bassani et al. 1999, $^n$ Landi et al. 2008, $^o$ Landi et al. 2007c, $^p$ Molina et al. 2008, $^q$ Sazonov
 et al. 2008, 
 $^s$ Landi et al. 2007a, $^t$ Gondoin et al. 2003, $^u$ Landi et al. 2007d, $^v$ Bassani et al. 2007, $^w$ Tomsick et
 al. 2008, $^x$ Ricci et al. 2009b

\begin{table}
\caption[]{Spectral fit of a cut-off power-law model for AGN with more complex IBIS/ISGRI
 spectra.}
\begin{tabular}{lcccc}
\hline\hline
Name & $N_{\rm H}$ & $\Gamma_{ISGRI}$ & $E_C$ & $\chi_\nu^2$ (d.o.f.)\\
    & $[10^{22} \rm \, cm^{-2}]$ & & $[\rm keV]$ & \\
\hline
Mrk 348  &  30 & $0.9 {+0.4 \atop -0.6}$ & $55 {+56 \atop -25}$ & 0.91
(8)\\
NGC 4151     &  $6.9$ & $1.60 {+0.06 \atop -0.07}$ & $118 {+21 \atop
  -13}$ & 1.10 (9)\\ 
NGC 4388     &  $27$ & $1.3 \pm 0.1$ & $95 {+26 \atop -17}$ & 0.34 (7) \\
NGC 4507     &  $29$ & $1.15 {+0.28 \atop -0.13}$ & $72 {+53 \atop
  -22}$ & 0.54 (8) \\
NGC 4593     &  0.02 & $1.1 {+0.4 \atop -0.2}$ & $48 {+114 \atop -9}$
& 0.87 (7) \\
NGC 4945     &  $400$ & $1.43 {+0.16 \atop -0.11}$ & $127{+56 \atop
  -26}$ & 0.75 (8) \\
4U 1344-60    &  $5$ & $1.11 {+0.29 \atop -0.09}$ & $51 {+27 \atop
  -12}$ & 0.76 (7) \\
IC 4329A      &  $0.42$ &   $1.37 \pm 0.17$ & $80 {+31 \atop -19}$ & 
1.50 (9)  \\
Circinus Galaxy&$360$ & $1.29 {+0.18 \atop -0.13}$ & $33 {+6 \atop
  -3}$ & 1.03 (7)  \\
GRS 1734-292 &  $3.7$ & $1.50 \pm 0.10$ & $54 {+8 \atop -7}$ & 0.88 (8) \\
IGR J17488-3253& $0.2$& $0.74 {+0.33 \atop -0.12}$ & $40 {+8  \atop
  -3}$ & 0.98 (7) \\
IGR J21247+5058 &$0.6$& $1.19 {+0.13 \atop -0.19}$ &$74 {+38  \atop
  -28}$ & 0.75 (8)\\

\hline
\end{tabular}
\label{complexISGRIspec}
\end{table}

\begin{longtable}{lrccccccc}
\caption{Spectral fit results for combined JEM-X and IBIS/ISGRI
  data\label{fitcombined}.}\\
\hline\hline
Name & JEM-X & $N_{\rm H}^b$ & $f_{3-20 \rm \, keV}^c$ & $f_{20-100 \rm \, keV}^c$ &
$\Gamma$ & $E_C$ & $\log L_{3 - 100 \rm \, keV}$& $\chi^2_\nu$
(d.o.f.)\\
 & $[\sigma]$ &  $[10^{22} \rm \, cm^{-2}]$ &  &   & & $[\rm keV]$ & $[\rm \, erg \, s^{-1}]$ &\\ 
\hline
\endfirsthead
\caption{continued.}\\
\hline\hline
Name & JEM-X & $N_{\rm H}$ & $f_{3-20 \rm \, keV}$ & $f_{20-100 \rm \, keV}$ &
$\Gamma$ & $E_C$ & $\log L_{3 - 100 \rm \, keV}$ & $\chi^2_\nu$
(d.o.f.)\\
 & $[\sigma]$ & $[10^{22} \rm \, cm^{-2}]$ &  &   & & $[\rm keV]$ & $[\rm \, erg \, s^{-1}]$ &\\ 
\hline
\endhead
\hline
\endfoot
1ES 0033+59.5   & 10.1 & $0.36$  & 6.0 & 1.4 & $2.75 {+0.18 \atop -0.16}$ &
-- & 45.16 & 0.51 (8)\\ 
NGC 1275$^a$    & 38.6 & $3.75$& 22.8 & 3.0 & -- & -- & 43.58 & $>5^a$\\
3C 120          & 5.5 & $0.2$ & 5.8 & 7.5& $1.76 \pm 0.06$ & -- &
44.52 & 0.67 (15) \\
Mrk 421         & 155.7 & $0.08$ & 61.6 & 32.5 & $2.12 {+0.05 \atop -0.04}$
& $133 {+32 \atop -21}$ & 45.29 & 1.43 (14)\\
NGC 4151$^a$        & 91.5 & $6.9$ & 33.2 & 40.4 & $1.75 {+0.04 \atop -0.06}$ &
$170 {+30 \atop -32}$ & 43.25 & 1.36 (14)\\
NGC 4388$^a$        & 14.3 & $27$ & 10.3 & 26.0 & $1.26 {+0.15 \atop -0.16}$ &
$78 {+35 \atop -20}$ & 43.68 & 0.41 (15)\\
NGC 4593 $^a$       & 7.0 & $0.02$  & 4.4 & 7.7 & $1.46 {+0.07 \atop -0.15}$ &
$193 {+12 \atop -93}$ & 43.34 & 1.01 (15) \\
Cen A$^a$           & 57.2 & $12.5$ & 28.1 & 43.4 & $1.85 \pm 0.01$ &
-- & 42.94 & 1.49 (17)\\
MCG-06-30-015   & 11.2 & $0.03$ & 7.0 & 5.0 & $2.10 \pm 0.05$
& -- & 42.93 & 0.79 (10) \\
4U 1344-60$^a$      & 8.3 & $5$ & 5.6 & 21.0 & $1.47 {+0.11 \atop -0.16}$ & $86
{+31 \atop -22}$ & 43.70 & 0.76 (15)\\
IC 4329A$^a$        & 11.3 & $0.42$ & 14.5 & 21.1 & $1.41 \pm 0.08$ &
$86 {+18 \atop -13}$ & 44.31 & 1.29 (15)\\
Circinus Galaxy$^a$ & 5.3 & $360$ & 3.9 & 21.0 & $1.2 \pm 0.2$ & $30 \pm
3$ & 42.03 & 1.52 (11)\\
NGC 5506        & 16.8 & $3.4$ & 12.1 & 8.0 & $2.20 \pm 0.08$ & -- &
43.27 & 0.66 (12)\\
H 1426+428      & 7.5 & $<0.02$ & 6.1 & 1.2 & $2.83 {+0.25 \atop -0.21}$ & --
& 44.90 & 0.67 (9)\\
Mrk 501         & 7.9 & $0.013$ & 10.2 & 4.8 & $2.3 \pm 0.1$ & -- &
44.60 & 0.73 (15)\\
GRS 1734-292$^a$    & 10.5 & $3.7$ & 6.7 & 8.0 & $1.25 {+0.10 \atop
  -0.16}$ & $39 {+5 \atop -6}$ & 44.02 & 1.28 (12)\\
IGR J17488-3253$^a$ & 21.8 & $0.72$ & 1.7 & 2.9 & $1.63 \pm 0.03$  &
--   & 43.11 & 1.10 (119)\\ 
QSO B1957+405$^a$   & 10.5 & $20$  & 16.2 & 8.3 & $1.9 {+0.1 \atop
  -0.2}$   & --   & 45.26 & 0.76 (9)\\
Mrk 509         & 7.2 & $<0.01$ & 7.5 & 9.3 & $1.8 \pm 0.1$&  --
& 44.66 & 0.69 (8)\\
IGR J21178+5139 & 17.1 & $2$ & 1.5 & 2.2 & $1.7 \pm 0.1$ & --  & -- &
1.61 (13) \\
IGR J21247+5058$^a$ & 54.1 & $0.6$ & 9.5 & 11.1 &  $1.4 \pm 0.1$ & $61
{+22 \atop -11}$       & 44.27 & 1.41 (11)\\
3C 454.3        & 6.4 & $0.5$ & 10.7 & 20.3 & $1.54 \pm 0.05$& --  &
47.93 & 0.58 (16)\\
1H 2251-179     & 10.7 & $<0.19$ & 5.7 & 7.2 & $1.8 \pm 0.1$  & --
& 45.10 & 1.20 (13) \\
\end{longtable}

$^a$ see Appendix \ref{section:singlesource},
$^b$ $N_H$ values have been fixed during the fit,
$^c$ fluxes are given in $[10^{-11},
\rm \, erg \, cm^{-2} \, s^{-1}]$
%


\onecolumn
\begin{longtable}{lccccrc}
\caption{Median V magnitude, the average of error estimates $\langle \sigma_V
\rangle$, luminosity in the Johnson V
filter,  \aox , the number of photometric points and contamination flag ("Y"
indicates potential contamination by a nearby star up to 0.2 mag; see text for
details).\\
\label{OMCresults}}\\
\hline\hline
Name   & $V$ [mag] & $\langle \sigma_V \rangle$ [mag] & $\log$~\lv\ [erg s$^{-1}$] & \aox & $N$& Cont. flag \\
\hline
\endfirsthead
\caption{continued.}\\
\hline\hline
Name   & $V$ [mag] & $\langle \sigma_V \rangle$ [mag] & $\log$~\lv\ [erg s$^{-1}$] & \aox & $N$ & Cont. flag \\
\hline
\endhead
\hline
\endfoot
          Mrk 348 &   13.76  &  0.05    & 42.71  & 1.01  &     81   &   Y  \\
          NGC 788 &   12.60  &  0.04    & 43.09  & 1.14  &   1342   &      \\
          NGC 985 &   13.73  &  0.05    & 43.66  & 1.13  &    644   &      \\ 
         NGC 1052 &   11.38  &  0.03    & 42.70  & 1.31  &   1004   &      \\
         NGC 1068 &    9.99  &  0.02    & 44.80  & 1.38  &    693   &      \\
         NGC 1275 &   12.51  &  0.04    & 43.27  & 1.21  &    856   &\\ 
         NGC 1365 &   11.51  &  0.05    & 42.72  & 1.27  &     35   &      \\
           3C 120 &   14.00  &  0.05    & 43.31  & 1.02  &    216   &      \\
         UGC 3142 &   15.12  &  0.22    & 43.49  & 0.93  &    259   &      \\
         ESO 33-2 &   13.99  &  0.09    & 42.78  & 1.09  &     40   &      \\
            Mrk 3 &   12.88  &  0.04    & 42.95  & 1.08  &    425   &   Y  \\
            Mrk 6 &   13.68  &  0.06    & 42.94  & 1.07  &    657   &      \\
    QSO B0716+714 &   14.27  &  0.09    & 45.26  & 1.19  &   1116   &      \\
       ESO 209-12 &   14.60  &  0.21    & 43.24  & 1.07  &    387   &      \\
      PG 0804+761 &   14.07  &  0.08    & 44.28  & 1.09  &    299   &      \\
       4U 0937-12 &   12.66  &  0.03    & 42.56  & 1.12  &    443   &      \\
    MCG-05-23-016 &   13.12  &  0.05    & 42.47  & 1.03  &   1514   &      \\
         NGC 3227 &   12.06  &  0.04    & 42.21  & 1.11  &     12   &      \\
         NGC 3281 &   12.67  &  0.04    & 42.85  & 1.16  &    316   &      \\
          Mrk 421 &   13.05  &  0.06    & 43.58  & 0.91  &   3112   &      \\
         NGC 4051 &   12.26  &  0.04    & 41.67  & 1.20  &     76   &      \\
         NGC 4138 &   11.87  &  0.04    & 42.06  & 1.26  &    585   &      \\
         NGC 4151 &   11.52  &  0.04    & 42.28  & 1.03  &   7461   &      \\
           Mrk 50 &   14.47  &  0.09    & 42.86  & 1.11  &    774   &   Y  \\
         NGC 4388 &   12.26  &  0.06    & 42.80  & 1.05  &    509   &      \\
         NGC 4395 &   14.09  &  0.11    & 40.23  & 1.05  &      7   &      \\
           3C 273 &   12.58  &  0.04    & 45.31  & 1.05  &   2010   &   Y  \\
  IGR J12391-1612 &   14.15  &  0.06    & 43.34  & 1.05  &    104   &      \\
         NGC 4593 &   12.24  &  0.04    & 42.83  & 1.14  &   1204   &      \\
       ESO 323-32 &   13.11  &  0.07    & 43.03  & 1.20  &    158   &  Y  \\
           3C 279 &   15.52  &  0.11    & 45.29  & 1.01  &    294   &      \\
          Mrk 783 &   15.39  &  0.16    & 43.39  & 1.01  &     37   &      \\
         NGC 4945 &   12.44  &  0.17    & 41.43  & 1.04  &    327   &      \\
  IGR J13091+1137 &   13.70  &  0.05    & 43.18  & 1.07  &     15   &      \\
         NGC 5033 &   11.57  &  0.03    & 42.16  & 1.33  &     55   &      \\
       ESO 383-18 &   14.33  &  0.08    & 42.32  & 1.09  &    673   &   Y  \\
    MCG-06-30-015 &   13.23  &  0.07    & 42.36  & 1.09  &   1268   &   Y  \\
          Mrk 268 &   14.30  &  0.06    & 43.37  & 1.08  &      7   &      \\
         IC 4329A &   13.20  &  0.05    & 43.00  & 0.98  &    501   & \\
         NGC 5506 &   12.93  &  0.05    & 42.38  & 1.03  &    701   &   \\
         NGC 5548 &   13.14  &  0.05    & 43.08  & 1.17  &    256   &      \\
       H 1426+428 &   16.11  & 0.25     & 43.72  & 0.98  &    619   &      \\
         NGC 5995 &   13.47  & 0.07     & 43.29  & 1.05  &    118   &   Y  \\
        ESO 138-1 &   13.59  & 0.11     & 42.34  & 1.20  &     12   &   Y  \\
         NGC 6221 &   11.91  & 0.07     & 42.49  & 1.33  &     16   &      \\
          Mrk 501 &   13.21  & 0.04     & 43.65  & 1.07  &    567   &      \\
         NGC 6300 &   11.52  & 0.05     & 42.38  & 1.19  &     31   &      \\
         3C 390.3 &   14.61  & 0.13     & 43.54  & 1.00  &    482   &   Y  \\
      1H 1934-063 &   13.35  & 0.05     & 42.59  & 1.10  &     12   &      \\
         NGC 6814 &   12.42  & 0.04     & 42.31  & 1.14  &    134   &      \\
       ESO 399-20 &   13.96  & 0.08     & 43.07  & 1.18  &     84   &      \\
          Mrk 509 &   13.38  & 0.05     & 43.60  & 1.03  &    400   &      \\
         NGC 7172 &   12.66  & 0.04     & 42.66  & 1.10  &    508   &      \\
          Mrk 915 &   14.10  & 0.09     & 42.75  & 1.09  &     40   &      \\
      1H 2251-179 &   14.20  & 0.09     & 43.83  & 0.98  &   1570   &   Y  \\
         NGC 7469 &   12.70  & 0.04     & 44.21  & 1.15  &     22   &      \\
   MCG-02-58-022  &   14.13  & 0.06     & 43.59  & 1.01  &     93   &      \\
\end{longtable}

\begin{table}
\caption[]{Average properties of the \textit{INTEGRAL} AGN. 
In parentheses, the number of objects used for the given average value
is indicated.}
\begin{tabular}{lcccc}
\noalign{\smallskip}
\hline\hline
   & Seyfert 1 & Seyfert 1.5 & Seyfert 2 \\
\hline
$\langle z \rangle$  & 0.03 (63) & 0.014 (24) & 0.02 (57) \\
$\langle \Gamma \rangle^a$  & 1.92 $\pm$ 0.02 & 2.02 $\pm$ 0.03
& 1.88 $\pm$ 0.02 \\
 & (55) & (20) & (44)\\
$\langle \log \, N_{\rm H} \rangle^b$  & 21.2 (61) & 21.7 (23) & 22.9 (51) \\
$\langle \log \, L_{\rm 20-100 \, keV}\rangle^c$  & 44.0 (63) & 43.3 (24) & 43.4 (57) \\
$\langle \log \, M_{\rm BH}\rangle$  & 7.8 (30) & 7.2 (14) & 7.7 (27) \\
$\langle \lambda \rangle^d$  & 0.064 (30) & 0.015 (14) & 0.02 (27) \\
\hline
   & Unabs &  Absorbed & all Sey \\
\hline
$\langle z \rangle$  & 0.03 (74) &     0.014 (60) &     0.02 (144) \\
$\langle \Gamma \rangle^a$  & 1.94 $\pm$ 0.02 & 1.91 $\pm$ 0.02 & 1.93 $\pm$ 0.01 \\
& (66) & (44) & (119)\\
$\langle \log \, N_{\rm H} \rangle^b$  & 21.0 (75) & 23.1 (60) & 21.9 (135) \\
$\langle \log \, L_{\rm 20-100 \, keV} \rangle^c$  & 43.8 (74) & 43.4 (60) & 43.6 (144) \\
$\langle \log \, M_{\rm BH}\rangle$  & 7.6 (37) & 7.7 (34) & 7.6 (71) \\
$\langle \lambda \rangle^d$  & 0.06 (37) & 0.015 (34) & 0.03 (71) \\
\hline
\end{tabular}

$^a$ $\Gamma$ is the average photon index derived
from the weighted mean on the power law model fits to the single
IBIS/ISGRI spectra.\\
$^b$ absorption is given in $[\rm cm^{-2}]$\\
$^c$ luminosities in $[\rm erg \, s^{-1}]$\\
$^d$ $\lambda = L_{Bol} / L_{Edd}$ is the Eddington ratio
\label{average}
\end{table}

\begin{table}
\caption[]{Results from spectral fitting of the stacked IBIS/ISGRI
 spectra of INTEGRAL AGN}
\begin{tabular}{lcccc}
\noalign{\smallskip}
\hline\hline
sample    & $\Gamma$ & $E_C \, [\rm keV]$ & $R$ & $\chi^2_\nu$\\
\hline
Sey~1 ($\ge 5\sigma$) & $1.96 {+0.03 \atop -0.02}$ & -- & -- & 5.66\\
                   & $1.44 \pm 0.10$ & $86 {+21 \atop -14}$ & -- &
                   1.10\\
\, \, \, \,$(i = 30^\circ)$   & $1.96 \pm 0.02$ & -- & $1.2 {+0.6 \atop -0.3}$ &
1.15\\
\hline
Sey~1.5 ($\ge 5\sigma$) & $2.02 \pm 0.04$ & -- & -- & 3.54\\
                   & $1.36 \pm 0.15$ & $63 {+20  \atop -12}$ & -- &
                   0.57\\
\,\, \, \,  $(i = 45^\circ)$   & $2.04 \pm 0.04$ & -- & $3.1 {+4.7 \atop -1.3}$
                   & 0.29\\
\hline
Sey~2 ($\ge 5\sigma$) & $1.89 {+0.04 \atop -0.02}$ & -- & -- & 3.13\\
                   & $1.65 \pm 0.05$ & $184 {+16 \atop -52}$ & -- &
                   2.58\\
\,\,  \, \, $(i = 60^\circ)$   & $1.91 {+0.02 \atop -0.03}$ & -- & $1.1  {+0.7 
                   \atop -0.4}$ & 1.67\\
\hline
all Sey ($\ge 5\sigma$) & $1.97 \pm 0.02$ & -- & -- & 6.18\\
                   & $1.44 {+0.08 \atop -0.13}$ & $86 {+16 \atop
                     -17}$ & -- & 1.96\\
\,\, \, \,  $(i = 45^\circ)$   & $1.95 \pm 0.02$ & -- & $1.3 {+0.7 \atop
                     -0.4}$ & 1.53\\
\hline
all Sey ($\ge 10\sigma$) & $1.98  \pm 0.02$ & -- & -- &
6.04\\
                   & $1.41 {+0.10 \atop -0.11}$ & $80 {+17 \atop
                     -13}$ & -- & 1.23\\
\,\, \, \,  $(i = 45^\circ)$   & $1.95 \pm 0.02$ & -- & $1.3 {+0.7 \atop
                     -0.4}$ & 1.23\\
\hline
Unabs. ($\ge 5\sigma$) & $1.97 {+0.03 \atop -0.01}$ & -- & -- & 5.64\\
                   & $1.53 {+0.09 \atop -0.08}$ & $100 {+25 \atop -15}$ & -- &                    1.82\\
\,\,  \, \,  $(i = 45^\circ)$    & $1.98 \pm 0.02$ & -- & $1.3 {+0.6 \atop -0.4}$ & 0.95\\ 
\hline
Abs. ($\ge 5\sigma$) & $1.91 {+0.04 \atop -0.03}$ & -- & -- & 1.2\\
                   & $1.43 {+0.13 \atop -0.08}$ & $ 94 {+32 \atop -13}$ & -- &                    1.59\\
\,\,  \, \, $(i = 45^\circ)$    &  $1.91 {+0.02 \atop -0.03}$ & -- & $1.5 {+1.5 \atop -1.4}$ & 1.03\\
\end{tabular}
\label{spectralfits}
\end{table}

\begin{table}
\caption[]{Correlation matrix for {\integral} AGN}
\begin{tabular}{lccccc}
\hline\hline
    & $M_{BH}$ & $\Gamma$ & $L_X$ &
    Edd. ratio & $L_V$\\
          &          &          &       & $\lambda$ & \\
\hline
$N_{\rm H}$    & no       & no       & no       & no   & no\\
$M_{BH}$ & --       & no       & $>99.99 \%$ & intrinsic   & $>99.99 \%$ \\
$\Gamma$ &          & --       & no     & no & no\\
$L_X$    &          &          & --     & intrinsic   & $>99.99 \%$ \\
$\lambda$&          &          &        & --   & no\\
\hline
\end{tabular}
\label{correlations}

\end{table}

\begin{table}
\caption[]{OSSE detected AGN not seen by {\integral}}
\begin{tabular}{lcc}
\hline\hline
Name & Type & exposure$^a$\\
    &      & [ks]      \\
\hline
CTA 102  & blazar & 180 \\
H 1517+656 & BL Lac& 19 \\
III Zw2 & Sy1/2 & 15 \\
Mrk 279 & Sy1.5 & 11 \\
M82 & Starburst & 127\\
NGC 253 & Starburst & 1 \\
NGC 7213 & Sy1.5 & 31 \\
NGC 7582 & Sy2 & -- \\
PKS 2155-304 & BL Lac & 347 \\
QSO 1028+313 & Sy1.5 & 556\\

\hline
\end{tabular}
\label{OSSEnonINTEGRAL}

$^{a}$ ISGRI exposure time\\
\end{table}

\appendix
\section{Notes on individual sources}
\label{section:singlesource}

We include here all Seyfert galaxies above $30 \sigma$ IBIS/ISGRI detection
significance, all sources showing a complex ISGRI spectrum (Table~\ref{complexISGRIspec}), and those
for which the
results found here differ from previous works.

Mrk 348: the X-ray spectrum of this Seyfert~2 was studied by
{\it RXTE}, showing the same spectral shape as reported here ($\Gamma
= 1.8$) and evidence for a reflection component with $R \lae 1$
(\cite{Smith01}). Instead of a cut-off power law with $\Gamma =0.9$
and $E_C = 55 \rm \, keV$ as given in Tab.~\ref{complexISGRIspec}, the
\integral 
data can be equally well represented by a Compton reflection model
(PEXRAV) with $R = 1$ and photon index $\Gamma =
1.8$ and no high-energy cut-off. 

NGC~1275 presents a very complex spectrum in \integral data as it
includes several components of different physical origin. While the
hard X-ray spectrum visible in the IBIS/ISGRI data is dominated by the
narrow-line radio galaxy NGC~1275 and its spectrum above 20 keV can be
  represented by a simple power law model, we observe in JEM-X the Perseus galaxy cluster. An extensive discussion of the \integral spectrum has been presented in Eckert \& Paltani (2009). 

NGC 4051: this Seyfert~1.5 shows a strong reflection component when fit together with soft X-rays,
e.g. 
$R \simeq 7$ for
{\it Suzaku} (\cite{Terashima09}), and $R \simeq 6$ for combined {\it
 Swift}/XRT and IBIS/ISGRI data (\cite{Beckmann09}), while the data
presented here allow only to fit a single power law model with $\Gamma
= 2.1 \pm 0.2$. When fitting the
{\it Suzaku} data with a simple power law model, Terashima et
al. (2009) derive a photon index of $\Gamma = 1.5 {+0.3 \atop -0.2}$
for a low flux state, indicating strong flux and spectral variability.

NGC 4151: this bright AGN allows complex modelling beyond the scope of
this paper, and we refer to an
early \integral analysis by Beckmann et al. 2005, to an analysis of
{\it BeppoSAX} data by de
Rosa et al. (2007), and to a study of the different spectral states by
Lubi\'nski et al. (2009).

NGC 4388: the hard X-ray data of this Seyfert~2 galaxy have been
studied by Beckmann et~al. 2004. Their analysis of \integral, {\it
 XMM-Newton}, {\it BeppoSAX}, {\it CGRO}, and {\it SIGMA} data showed
that the hard X-rays spectrum is well described by an absorbed power
law with $\Gamma = 1.65 \pm 0.04$ and $N_{\rm H} = 2.7 \times 10^{23}
\rm \, cm^{-2}$, with no indication of a cut-off or reflection
component. The data presented here now show evidence for a cut-off at
80 keV and $\Gamma = 1.3$. Recently, a turn over at $E_C = 30 \pm 13 \rm \, keV$ with
$\Gamma = 0.9 \pm 0.3$ was also reported in {\it Suzaku} data of
NGC~4388 (\cite{Shirai08}), detecting also significant spectral variability.

NGC 4507: this Seyfert~2 shows a reflection component of the order of
$R = 0.5 - 1$ in {\it BeppoSAX} observations and a photon index of
$\Gamma = 1.3 - 1.9$, while the cut-off energy was not constrained
(Dadina 2007). 
The IBIS/ISGRI data do not require the presence of reflection, and a
simple cut-off power law ($\Gamma=1.1 \pm  0.2$, $E_{\rm
  C}=65^{+27}_{-12}$ keV) is sufficient ($\chi^{2}=3.5$ for 7
dof). Applying the PEXRAV model without cut-off, we obtained a chi-squared of $\chi^{2}=6.2$ for 7 dof. The value of the photon index and the reflection obtained are
$\Gamma=1.7 \pm 0.1$ and $R=0.6^{+1.5}_{-0.5}$, respectively,
consistent with the {\it BeppoSAX} observations.

NGC 4593: {\it BeppoSAX} data of this Seyfert~1 showed a reflection
component with $R = 1.1 {+2.6 \atop -0.5}$ and $\Gamma = 1.9 \pm 0.1$ but no evidence for a
cut-off (\cite{Dadina07}). From IBIS/ISGRI and JEM-X data we found
that a power law with a cut-off gives a good fit to the data
($\Gamma=1.5 \pm 0.1$, $E_{\rm C}=193^{+12}_{-93}$ keV, $\chi
^{2}=15$ for 15 dof). The value of the absorption has been fixed to
$N_{\rm H} = 2 \times 10^{20} \rm \, cm^{-2}$. Applying a PEXRAV model
with no cut-off also provides a good representation of the data with
$\chi^2 = 10.7$ for 15 dof and gives a photon index of
$\Gamma=1.9 \pm 0.1$ and a reflection component of
$R=1.7^{+1.6}_{-0.8}$, both parameters consistent with the results by
Dadina et al. (2007). 

NGC 4945: the IBIS/ISGRI spectrum of the Seyfert 2 galaxy NGC~4945 (with an exposure of 276 ks) is amongst the six analyzed by Soldi et al. (2005). In their work the best model to the data is a simple power law ($\Gamma=1.9^{+0.1}_{-0.1}$), and they give a lower limit to the possible high energy cut-off, $E_{\rm C} \gg 130$ keV. Using the new IBIS/ISGRI and JEM-X data we found that a simple absorbed power law does not provide a good fit ($\chi ^{2}=30$ for 12 dof) and a cutoff at high energy is necessary. This component improves significantly the goodness of the fit ($\chi ^{2}=3.3$ for 11 dof) and gives a photon index of $\Gamma=1.4^{+0.2}_{-0.2}$ and a cutoff at $E_{\rm C}=121^{+63}_{-34}$ keV. The absorption has been fixed to $N_{\rm H} = 400 \times 10^{22} \rm \, cm^{-2}$. These values are also consistent with those obtained by Guainazzi et al. (2000), using {\it BeppoSAX} data.

Cen~A is the only extragalactic object also seen by IBIS/PICsIT. Combined
IBIS/ISGRI, SPI, and PICsIT data analysis gave a spectral slope of
$\Gamma = 1.80 \pm 0.01$ (\cite{PICsIT}), close to the results
presented here. It has been also detected by {\it Fermi}/LAT (\cite{FermiLAT}) and at very high energy gamma-rays by HESS (\cite{CenA_HESS}).

\begin{figure}
\includegraphics[height=8.5cm,angle=270]{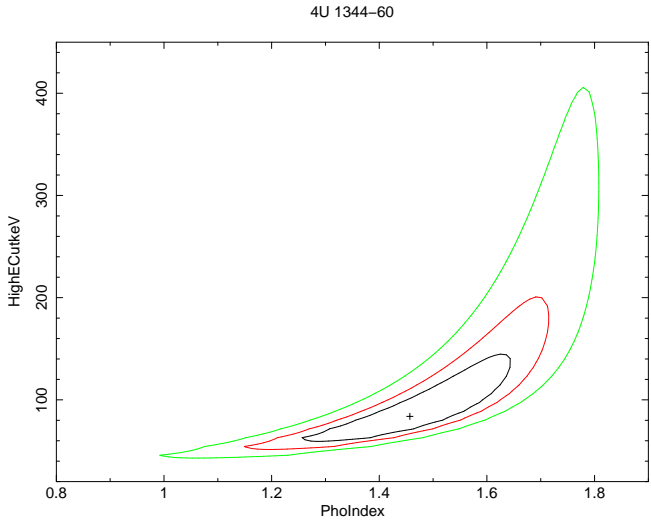}
\caption[]{Contour plot of the 1, 2, and 3 sigma confidence levels of
 cut-off energy $[\rm keV]$ versus photon index $\Gamma$. A steeper
 spectral fit requires a higher cut-off energy.}

\label{fig:4U1344_contour}
\end{figure}

4U 1344-60: 
in Beckmann et~al. (2006) the analysis of combined {\it XMM-Newton} and
{\it INTEGRAL} data showed a good representation of the broad-band data with an absorbed
power-law ($\Gamma = 1.65$) plus a Gaussian component. Applying the
same simple model to the IBIS/ISGRI data used here, leads to a
steep hard X-ray power law ($\Gamma = 1.9 \pm 0.1$) but gives a bad
fit result ($\chi^2_\nu > 2$). Using the simultaneous JEM-X and ISGRI
data and adding a cut-off, flattens the
spectrum: the lower the cut-off energy, the flatter the resulting
photon index ($\Gamma = 1.5 {+0.1 \atop -0.2}$, $E_C = 86 {+31 \atop
 -22}$), as can be seen in Fig~\ref{fig:4U1344_contour}. 
Panessa et~al. (2008) analysed again non-simultaneous {\it XMM-Newton} and
{\it INTEGRAL} data, finding a cut-off power law with photon index $\Gamma =
1.75 {+0.18 \atop -0.14}$ and $E_C > 78 \rm \, keV$. Fixing the
spectral slope for the combined JEM-X and IBIS/ISGRI spectrum to $\Gamma = 1.75$, we also
get a higher cut-off energy of $E_C = 214 {+73 \atop -44} \rm \,
keV$. 


IC 4329A: the {\it BeppoSAX} spectrum of this Seyfert~1 analysed by Dadina
(2007) by applying a PEXRAV model showed in several observations a cut-off at
energies $E_C > 120 \rm \, keV$, with a reflection component of $R =
0.5 - 1.5$ and a photon index of $\Gamma = 1.9 - 2.0$. Using the same
model for the combined JEM-X and ISGRI data, we get a 
flatter spectrum, with $\Gamma = 1.6 \pm 0.3$, $E_C = 124 {+243 \atop
 -48} \rm \, keV$ and $R = 0.3 {+0.8 \atop -0.3}$, and thus a
spectrum which is consistent with no reflection component. We
therefore applied the simpler cut-off power law model which leads to
$\Gamma = 1.4 \pm 0.1$ and $E_C = 86 {+18 \atop -13} \rm \, keV$. 

Circinus Galaxy: the combined IBIS/ISGRI and SPI spectrum of the Circinus galaxy
has been studied by Soldi et al. (2005)
based on 589 ks exposure time, finding as the best model to the data an absorbed power law
($\Gamma=1.8^{+0.4}_{-0.5}$) with $N_{\rm H} = 400 \times 10^{22} \rm \, cm^{-2}$ and a
high-energy cutoff at $E_{\rm C}=50^{+51}_{-18}$ keV, values
consistent with the ones obtained by previous {\it BeppoSAX}
observations. Using the same model and fixing the hydrogen column
density to the one used by Soldi et al. (2005), we found the best fit
to the data ($\chi^{2}=16$ for 11 dof), with parameters consistent
with the values listed above ($\Gamma=1.2 \pm 0.2$, $E_{\rm C}=30 \pm 3$ keV).
Dadina (2007) analysed two {\it BeppoSAX} observations applying
reflection models with
$\Gamma_1 = 1.7 \pm 0.1$, $R_1 = 0.29 {+0.05 \atop -0.04}$ and
$\Gamma_2 = 1.3 \pm 0.2$, $R_2 = 0.35 {+0.39 \atop -0.09}$, both with
cut-off at $E_C \lae 50 \rm \, keV$. Adding a reflection component to
the data presented here improves significantly the fit ($ \chi^{2}=6.1$ for 9 dof) but the parameter are not well constrained: $ \Gamma=1.8^{+0.4}_{-0.6}$, $E_{\rm C}=76^{+180}_{-30}$ keV and  $R=1.6^{+8.0}_{-1.1}$. 


PG~1416-129: the Seyfert~1 galaxy had been included in the first \integral AGN catalogue (\cite{BE06}). Subsequent analysis of the data with improved software showed that the detection of this source was indeed spurious, and we now do not consider this source to be an \integral detected object. 

IGR~J16351-5806: this Seyfert~2 galaxy has recently been claimed
to be a Compton thick AGN with $N_{\rm H} > 1.5 \times 10^{24} \rm \,
cm^{-2}$ (\cite{Malizia09}). This is based on the observation that the
hard X-ray spectrum is rather flat, as also shown here with $\Gamma =
1.5$, and that a strong iron K$\alpha$ line with $EW > 1 \rm \, keV$
is an indicator for a significant reflection component. Malizia et al. fit the spectrum therefore as a pure reflection spectrum. As the ISGRI data alone do not allow fitting the complex model, we assumed the simple model of an unabsorbed power law, as also observed in {\it Swift}/XRT data (\cite{Landi07a}).

IGR~J16426+6536: this narrow-line Seyfert~1 ($z = 0.323 \pm
 0.001$, Parisi et al. 2008, Butler et al. 2009) shows a rather high luminosity of $L_{(20 - 100 \rm \,
 keV)} = 2 \times 10^{46}
\rm \, erg \, s^{-1}$ but moderate mass of the central black hole ($M
 = 1.1 \times 10^7 \, M_\odot$). This leads to a large Eddington ratio of
 about $\lambda \simeq 90$ (defined in Sect.~\ref{NHcorrelations}). It has to be pointed out though, that the
 soft X-ray counterpart as detected by {\it XMM-Newton} exhibits
 only a flux of $1.1 \times 10^{-12} \rm \, erg \, cm^{-2} \,
 s^{-1}$ (Ibarra et al. 2008), about 10 times below what would be expected from the {\integral}
 detection with $f_{(20 - 40 \rm \, keV)} = 2.6 \times 10^{-11} \rm \, erg \, cm^{-2} \,
 s^{-1}$. This might indicate that this source is strongly absorbed,
 highly variable, or that the optical counterpart is a
 misidentification, leading to the high super-Eddington accretion
 rate. 

GRS 1734-292: the IBIS/ISGRI data of GRS~1734-292 (for a total
exposure of 4040 ks) have been analyzed by Molina et~al. (2006), along with {\it ASCA}/GIS data, who found that the best fit is obtained using an absorbed cut-off power law. Using the new IBIS/ISGRI and JEM-X data we found that the best model ($\chi^{2}=12.1$ for 11 dof) is an absorbed power-law with a cut-off and a reflection component with the following parameters: $N_{\rm H} = 3.7 \times 10^{22} \rm \, cm^{-2}$, $E_{\rm C}=220^{+200}_{-150}$ keV, $\Gamma=2.0^{+0.3}_{-0.4}$ and $R=2.7^{+3.1}_{-1.8}$.
The black hole mass of GRS~1734-292 has been provided by I. Papadakis
(private communication) based on the empirical relation found by Tremaine et al. (2002) between
the black hole mass and the stellar velocity dispersion $\sigma_s$, estimating $\sigma_s$ from the width of the [O III] line reported
by Marti et al. (1998).

IGR~J17488-3253: for this Seyfert~1 galaxy  ($z=0.02$) a fit of simultaneous JEM-X and IBIS/ISGRI data by a
 cut-off power law model results in  $\chi^2_\nu
= 5.8$ for 11 degrees of freedom. As the source is located in a dense
area, it is possible that the JEM-X data are contaminated by sources
within the field of view, especially as the JEM-X data are well
represented by a black body model with a temperature of $0.7 \rm \,
keV$. We therefore analysed {\it Swift}/XRT data of the source. A combined fit of XRT and ISGRI data results in $\chi^2_\nu = 1.1$ for 119 degrees of freedom, showing an absorbed power law model with $N_{\rm H} = 0.72 \pm 0.04$ and $\Gamma = 1.63 \pm 0.03$, representing the data well over the $0.3 \rm \, keV$ to $100 \rm \, keV$ energy range. 


QSO B1957+405 (Cyg A): the IBIS/ISGRI (together with {\it
 BeppoSAX}/MECS and PDS) spectrum of Cygnus~A has been analyzed by Molina et al. (2006), using data for a total exposure of 426 ks. In their work they fitted the high-energy spectrum with a complex model in order to take into account also the gas emission of the galaxy cluster to which the AGN belongs. For doing so they used an absorbed power law plus a bremsstrahlung component. Following their work we fitted the IBIS/ISGRI and JEM-X spectrum using the same model (plus a cross-calibration constant) and we found that it provides the best fit ($\chi ^{2}=6$ for 9 dof) and that the parameters obtained ($\Gamma=1.9^{+0.1}_{-0.2}$ and $kT=4.5^{+3.7}_{-2.7}$ keV) are in good agreement with those they obtained. The value of the absorption has been fixed to $N_{\rm H} = 2 \times 10^{23} \rm \, cm^{-2}$.

IGR~J21247+5058: the high-energy broad-band spectrum of this radio galaxy was obtained by Molina et~al. (2007) by combining
{\it XMM-Newton} and {\it Swift}/XRT observation with
IBIS/ISGRI data. The
0.4--100 keV spectrum is well described by a power law, with slope $\Gamma =
1.5$, characterized by complex absorption due to two layers of material
partially covering the source and a high-energy cut-off around 70--80
keV, consistent with our findings ($\Gamma = 1.4 \pm 0.1$, $E_C = 61
{+22 \atop -11} \rm \, keV$) using JEM-X and ISGRI data. As Molina et
al. point out, features such as a narrow iron line and a Compton reflection
component, if present, are weak, suggesting that reprocessing of the
power-law photons in the accretion disc plays a negligible role in the
source.

\end{document}